\documentclass[12pt]{article}
\pdfoutput=1
\usepackage{amsmath, dsfont, amssymb, amscd, caption,  color}
\usepackage{graphicx,subfigure}
\usepackage[margin=1in]{geometry}

\usepackage{times,soul}
\usepackage{natbib}

\usepackage{setspace}
\usepackage{tikz}
\usepackage{url}

\usepackage{amsfonts}
\usepackage{fancybox}
\usepackage{algorithmic}
\usepackage{algorithm}
\usepackage{comment}
\usepackage{color}
\usepackage[colorlinks,citecolor=blue,urlcolor=blue]{hyperref}

\usepackage{paralist}
\usepackage{tikz}
\usetikzlibrary{arrows,shapes,snakes,automata,backgrounds,petri}

\newcommand\bm[1]{\boldsymbol{#1}}

\newcommand{\bpm}{\begin{pmatrix}}
\newcommand{\epm}{\end{pmatrix}}

\newcommand{\constantFiveP}{ c_3 }
\newcommand{\constantSixP}{ c_5 }
\newcommand{\constantSevenP}{ c_4 }

\newcommand{\constantFivePrime}{ c'_3 } 
\newcommand{\constantFivePP}{ c_3^{''} } 
\newcommand{\constantEight}{ c_6 } 

\newcommand{\constantTenP}{c_{1}}
\newcommand{\constantElevenP}{c_{2}}

\newcommand{\constantTwelveP}{a_{1}}
\newcommand{\constantThirteenP}{a_{2}}
\newcommand{\constantFourteenP}{a_{3}}
\newcommand{\constantFifteenP}{a_{4}}

\newcommand{\constantSixteen}{a_{5}}
\newcommand{\constantSeventeen}{a_{6}}
\newcommand{\constantEightteen}{a_{7}}

\newcommand{\constantSixteenP}{a'_{5}}

\newtheorem{theorem}{Theorem}
\newtheorem{lemma}{Lemma}

\newtheorem{corollary}{Corollary}
\newtheorem{proposition}{Proposition}
\newtheorem{assumption}{Assumption}

\newtheorem{remark}{Remark}
\newtheorem{example}{Example}

\def\T{{ \mathrm{\scriptscriptstyle T} }}

\newcommand*{\medcap}{\mathbin{\scalebox{1.5}{\ensuremath{\cap}}}}%

\newcommand{\indic}{\mathds{1}}

\newcommand*{\QEDB}{\hfill\ensuremath{\square}}%

\begin{document}

\title{The Multivariate Hawkes Process in High~Dimensions: Beyond Mutual Excitation}
\author{Shizhe Chen, Ali Shojaie, Eric Shea-Brown, and Daniela Witten  \\
}

\date{June 18th, 2019}

\maketitle

\begin{abstract}
The Hawkes process is a class of point processes whose future depends on their own history.
Previous theoretical work on the Hawkes process is limited to a special case in which a past event can only increase the occurrence of future events, and the link function is linear.
However, in neuronal networks and other real-world applications, inhibitory relationships may be present, and the link function may be non-linear.
In this paper, we develop a new approach for investigating the properties of the Hawkes process without the restriction to mutual excitation or linear link functions.
To this end, we employ a thinning process representation and a coupling construction to bound the dependence coefficient of the Hawkes process.
Using recent developments on weakly dependent sequences, we establish a concentration inequality for second-order statistics of the Hawkes process.
We apply this concentration inequality to cross-covariance analysis in the high-dimensional regime, and we verify the theoretical claims with simulation studies.
\end{abstract}


 \section{Introduction}\label{sec::intro}

 \cite{hawkes1971} proposed a class of point process models in which a past event can affect the probability of future events.
 The Hawkes process and its variants have been widely applied to model recurrent events in many fields, with notable applications to earthquakes  \citep{ogata1988}, crimes \citep{mohler2011}, interactions in social networks \citep{simma2012,perry2013}, financial events \citep{chavez2005,bowsher2007,sahalia2015}, and spiking histories of neurons (see e.g.,  \citealp{brillinger1988, okatan2005, pillow2008}).

 There is currently a significant gap between applications and statistical theory for the Hawkes process.
 \cite{hawkes1971} considered the \emph{mutually-exciting} Hawkes process, in which an event \emph{excites} the process, i.e. one event may trigger future events.
 Later, \cite{hawkes1974} developed a cluster process representation for the mutually-exciting Hawkes process, which is an essential tool for subsequent theoretical developments \citep{reynaud2010, hansen2015,bacry2015}.
 The cluster process representation requires two key assumptions: (i) the process is mutually-exciting;  and (ii) the link function is linear, implying that the effects of past events on the future firing probabilities are additive.
 In many applications, however, one might wish to allow for inhibitory events and non-additive aggregation of effects from past events.
 For instance, it is well-known that a spike of one neuron may \emph{inhibit} the activities of other neurons (see e.g., \citealp{purves2001}), meaning that it decreases the probability that other neurons will spike.
 Furthermore, non-linear link functions are often used when analyzing spike trains \citep{paninski2007,pillow2008}.
 In these cases, many existing theoretical results do not apply, since \citeauthor{hawkes1974}'s cluster process representation is no longer viable.

 In this paper, we propose a new analytical tool for the  Hawkes process that applies beyond the mutually-exciting and linear setting.
 We employ a new representation of the Hawkes process to replace the  cluster process representation.
 To demonstrate the application of this new analytical tool, we establish a concentration inequality for second-order statistics of the Hawkes process,  without restricting the process to be mutually-exciting or to have a linear link function. We apply this tool to study smoothing estimators of cross-covariance functions of the Hawkes process.

 While this paper was under revision, it came to our attention that \cite{costa2018} have concurrently studied the Hawkes process with inhibitions in the one-dimensional case. By contrast, our work allows for multiple dimensions and considers the  high-dimensional setting. We will provide some additional remarks on their proposal in Section~\ref{sec::highd}.

 The paper is organized as follows.
 We introduce the Hawkes process and review the existing literature in Section~\ref{sec::model}.
 In Section~\ref{sec::conc}, we present the construction of a coupling process, and derive a new concentration inequality for the Hawkes process.
 In Section~\ref{sec::cross-covariance}, we study the theoretical properties of smoothing estimators of the cross-covariance functions of the Hawkes process, and corroborate our findings with numerical experiments.
 Proofs of the main results are in Section~\ref{sec::proofs_conc}.
 We conclude with a discussion in Section~\ref{sec::discussion}.
 Technical proofs are provided in the Appendix.

 \section{Background on the  Hawkes Process}\label{sec::model}

 In this section, we provide a very brief review of point processes in general, and the Hawkes process in particular.
 We refer interested readers to the monograph by \cite{daley2003} for a comprehensive discussion of point processes.

 We use the notation $f\ast g (t) \equiv \int_{-\infty}^{\infty} f(\Delta) g(t-\Delta) \mathrm{d} \Delta $ to denote the convolution of two functions, $f$ and $g$.
 We use $\|a\|_2$ to denote the $\ell_2$-norm of a vector $a \in \mathbb{R}^p$.
 Furthermore, $\|f \|_{2,[l,u]} \equiv  \big\{\int_l^u f^2(t) \mathrm{d} t \big\}^{1/2}$ will denote the $\ell_2$-norm of a function $f$ on the interval $[l,u]$, and $\|f\|_{\infty} \equiv \sup_x |f(x)|$ will denote the maximum of $f$.
 We use $\Gamma_{\max}(\bm{A})$ for the maximum eigenvalue of a square matrix $\bm{A}$.
 We use $\bm{J}$ to denote a $p$-vector of ones.
 The notation $\indic_{[C]}$ is an indicator variable for the event $C$.

 \subsection{A Brief Review of Point Processes}

 Let $\mathcal{B}(\mathbb{R})$ denote the Borel $\sigma$-field of the real line,
 and let $\{t_{i}\}_{i \in\mathbb{Z}}$ be a sequence of real-valued random variables such that $t_{i+1}> t_{i}$ and $0 \leq t_1$.
 Here, time $t=0$ is a reference point in time, e.g., the start of an experiment.
 We define a simple point process $N$ on $\mathbb{R}$ as a family $\{N(A) \}_{A \in \mathcal{B}(\mathbb{R})}$ that takes on non-negative integer values such that the sequence $\{t_{i}\}_{i \in\mathbb{Z}}$ consists of event times of the process $N$, i.e., $N(A) = \sum_i \indic_{[ t_{i} \in A]}$ for $A \in \mathcal{B}(\mathbb{R})$.
 We write ${N}\big([t,t+\mathrm{d} t) \big)$ as $\mathrm{d} N_j(t)$, where $\mathrm{d} t$ denotes an arbitrarily small increment of $t$.

 Now suppose that $N$ is a \emph{marked} point process,  in the sense that  each event time $t_i$ is associated with a mark $m_i \in  \{1,\ldots, p\}$ \citep[see e.g., Definition 6.4.I. in][]{daley2003}.
 With a slight abuse of notation, we can then view $N$ as a multivariate point process, $\bm{N} \equiv \big( N_j\big)_{j=1,\ldots, p}$,
  for which the $j$th component process, $N_j$, satisfies $N_j(A) = \sum_i \indic_{[ t_{i} \in A, m_i = j]}$ for $A \in \mathcal{B}(\mathbb{R})$.
 To simplify the notation, in what follows, we will let   $\{t_{j,1},t_{j,2},\ldots\}$ denote the event times of $N_j$.
 Let $\mathcal{H}_t$ denote the history of $\bm{N}$ up to time $t$.
 The \emph{intensity process} $\bm{\lambda}(t) =\big( \lambda_1(t), \ldots, \lambda_p(t) \big)^{\T}$ is a $p$-variate $\mathcal{H}_t$-predictable process, defined as
  \begin{equation}\label{eqn::intensity_definition}
  \lambda_j(t) \mathrm{d} t =\mathbb{P}( \mathrm{d}  N_j(t) =1 \mid \mathcal{H}_{t}), \ j=1,\ldots, p.
 \end{equation}

 \subsection{A Brief Overview of the Hawkes Process}\label{sec::hawkes_rev}

 For the Hawkes process \citep{hawkes1971}, the intensity function \eqref{eqn::intensity_definition} takes the form
 \begin{equation}\label{eqn::HP_intensity_general}
 \lambda_{j}(t)= \phi_j \left\{\mu_{j} + \sum_{k=1}^p \big( {\omega}_{k,j} \ast  \mathrm{d} {N}_k \big) (t) \right\}, \;\;\; j=1,\ldots,p,
 \end{equation}
 where
 $$\big( {\omega}_{k,j} \ast  \mathrm{d} {N}_k \big) (t) = \int_0^{\infty} \omega_{k,j}(\Delta) \mathrm{d} N_k(t-\Delta)  = \sum_{i:t_{k,i} \leq t} \omega_{k,j}(t-t_{k,i}). $$
  We refer to $\mu_{j} \in \mathbb{R}$ as the \textit{background intensity}, and $\omega_{k,j}(\cdot): \mathbb{R}^{+} \mapsto \mathbb{R}$ as the \textit{transfer function}.
 If the \emph{link function} $\phi_j$ on the right-hand side of \eqref{eqn::HP_intensity_general} is non-linear, then $\lambda_j(t)$ is  the intensity of a non-linear Hawkes process \citep{bremaud1996}.
  We will refer to the class of  Hawkes processes that allows for non-linear link functions and negative transfer functions as the \emph{generalized Hawkes process}.

 In this paper, we assume that we observe the event times of a \emph{stationary} process $\mathrm{d}\bm{N}$ on $[0, T]$, whose intensities follow \eqref{eqn::HP_intensity_general}.
 The existence of a stationary process is guaranteed by the following assumption.
 \begin{assumption}
 	We assume that $\phi_j(\cdot)$ is $\alpha_j$-Lipschitz for $j=1,\ldots, p$.
 Let $\bm{\Omega}$ be a $p \times p$ matrix whose entries are  $\Omega_{j,k} =\alpha_j \int_0^{\infty} |\omega_{k,j}(\Delta)| \, \mathrm{d}\Delta$ for $1 \leq j,k \leq p$. We assume that  there exists a generic constant $\gamma_{\Omega}$ such that $\Gamma_{\max}(\bm{\Omega})\leq \gamma_{\Omega} < 1$.  \label{asmp::spectralradius}
 \end{assumption}
 Note that in Assumption~\ref{asmp::spectralradius}, the constant $\gamma_{\Omega}$ does not depend on the dimension $p$.
 For any fixed $p$, \cite{bremaud1996} establish that the intensity process of the form \eqref{eqn::HP_intensity_general} is stable in distribution, and thus a stationary process $\mathrm{d}\bm{N}$ exists given Assumption~\ref{asmp::spectralradius}.
 We refer interested readers to \cite{bremaud1996} for a rigorous discussion of stability for the Hawkes process.

 We define the \emph{mean intensity} $\bm{\Lambda}=(\Lambda_1, \ldots, \Lambda_p)^{\T} \in \mathbb{R}^p$  as
 \begin{equation}\label{eqn::Lambda}
 {\Lambda}_j \equiv \mathbb{E} [ \mathrm{d}  {N}_j(t)] /\mathrm{d} t, \ j=1,\ldots, p.
 \end{equation}
 Following  Equation 5 of \cite{hawkes1971}, we define
 the \emph{(infinitesimal) cross-covariance}  $\bm{V}(\cdot)  =\big( V_{k,j}(\cdot) \big)_{p \times p}: \mathbb{R} \mapsto \mathbb{R}^{p \times p}$ as
 \begin{equation}\label{eqn::V}
 {V}_{k,j}(\Delta )  \equiv \begin{cases}
 \mathbb{E}[\mathrm{d}  N_j(t) \mathrm{d}  N_{k}(t-\Delta )] /\{\mathrm{d}  t \mathrm{d}  (t-\Delta)\} -  \Lambda_j \Lambda_k & j \neq k\\
 \mathbb{E}[\mathrm{d}  N_k(t) \mathrm{d}  N_{k}(t-\Delta )] /\{\mathrm{d}  t \mathrm{d}  (t-\Delta)\} - \Lambda_k^2 - \Lambda_k   \delta(\Delta) & j = k\\
 \end{cases},
 \end{equation}
 for  any $\Delta \in \mathbb{R} $,  and $1\leq j,k \leq p$.
 Here $\delta(\cdot)$ is the Dirac delta function, which satisfies $\delta(x)=0$ for $x \neq 0$ and $\int_{-\infty}^{\infty}\delta(x) \mathrm{d} x =1$.

 \begin{example} (Linear Hawkes processes)
 Suppose that $\phi(x)=x$ and $\omega_{k,j}(\Delta) \geq 0$ for all $\Delta \in \mathbb{R}^{+}$ and for all $j,k \in \{1,\ldots, p\}$.
 The intensity in \eqref{eqn::HP_intensity_general} takes the form
 \begin{equation}\label{eqn::HP_intensity}
 \lambda_{j}(t)= \mu_{j} + \sum_{k=1}^p \big( {\omega}_{k,j} \ast  \mathrm{d} {N}_k \big) (t), \;\;\; j=1,\ldots,p.
 \end{equation}
 This is known as the linear Hawkes process \citep{hawkes1971,bremaud1996,hansen2015}. If the Hawkes process defined in \eqref{eqn::HP_intensity} is stationary, then the following relationships hold between the mean intensity $\bm{\Lambda}$ \eqref{eqn::Lambda}, the cross-covariance $\bm{V}$ \eqref{eqn::V}, the background intensity $\bm{\mu}\equiv (\mu_1,\ldots, \mu_p)^{\T}$ \eqref{eqn::HP_intensity}, and  the transfer functions $\bm{\omega} \equiv ( \omega_{k,j} )_{p \times p}$ \eqref{eqn::HP_intensity} (see, e.g., Equations 21 and 22 in \citealp{hawkes1971} or Theorem~1 in \citealp{bacry2014}):
 	\begin{equation}\label{eqn::wheq_I}
 	\bm{\Lambda}= \bm{\mu} + \left[ \int_0^{\infty} \bm{\omega} (\Delta) \mathrm{d} \Delta \right] \bm{\Lambda},
 	\end{equation}
 	and
 	\begin{equation}\label{eqn::wheq_II}
 	\bm{V}(\Delta)= \bm{\omega}(\Delta) {\rm diag}(\bm{\Lambda} ) + (\bm{\omega} * \bm{V})( \Delta),
 	\end{equation}
 	where $[\bm{\omega} * \bm{V}]_{k,j} (\Delta) \equiv \sum_{i=1}^p [\omega_{k,i}*V_{i,j}] (\Delta)$.
 Equation \eqref{eqn::wheq_II} belongs to a class of integral equations known as the Wiener-Hopf integral equations.
 For the linear Hawkes process, these equations are often used to learn the transfer functions $\bm{\omega}$ by plugging in estimators of $\widehat{\bm{\Lambda}}$ and $\widehat{\bm{V}}$ (see, e.g., \citealp{bacry2014}, \citealp{krumin2010}).
 These equations are similar to the Yule-Walker equations in  the vector-auto regression model \citep{yule1927,walker1931}.
 	\end{example}

 	\begin{remark}
 	An assumption similar to Assumption~\ref{asmp::spectralradius}, known as walk summability, was proposed by \cite{anandkumar2012} in the context of the Gaussian graphical model.
 	Consider a linear Hawkes process \eqref{eqn::HP_intensity} with $\omega_{k,j}(\Delta) \geq 0$ for all $\Delta$, so that $\bm{\Omega} = \int_0^{\infty} \bm{\omega}(\Delta) \mathrm{d} \Delta$.
 	Under Assumption~\ref{asmp::spectralradius}, we can rewrite \eqref{eqn::wheq_I} as
 	\begin{equation}\label{eqn::infinity_Lambda}
 	\bm{\Lambda} = \sum_{i=0}^{\infty} \bm{\Omega}^i \bm{\mu},
 	\end{equation}
 	where $\bm{\Omega}^i$ is the $i$th power of the matrix $\bm{\Omega}$.
 	In \eqref{eqn::infinity_Lambda}, $\bm{\Omega}^i \bm{\mu}$ can be seen as the intensity induced through paths of length $i$.
 	Assumption~\ref{asmp::spectralradius} ensures that  the induced intensity decreases exponentially fast as the path length $i$ grows, i.e., $\|\bm{\Omega}^i \bm{\mu}\|_2 \leq \gamma_{\Omega}^i \|\bm{\mu}\|_2$.
 	Viewed another way, the equality in \eqref{eqn::infinity_Lambda} says that the mean intensity is a sum of induced intensities through paths of all possible lengths,
 	and Assumption~\ref{asmp::spectralradius} prevents this sum from diverging. 
 	\end{remark}

  \section{A New Approach for Analyzing the Hawkes Process}\label{sec::conc}

  In this section, we present a new approach for analyzing the statistical properties of the Hawkes process, without assuming linearity of the link function $\phi_j$ or nonnegativity of the transfer function $\omega_{k,j}$ in \eqref{eqn::HP_intensity_general}.
  We provide an overview of existing theoretical tools and a new approach for analyzing the Hawkes process in Section~\ref{sec::overview_framework}.
  In Section~\ref{sec::coupling_framework}, we construct a coupling process using the \emph{thinning process representation}.
  In Section~\ref{sec::highd}, we present a bound on the \emph{weak dependence coefficient} for the Hawkes process using the coupling technique \citep{dedecker2004}, and present a new concentration inequality for the Hawkes process in the high-dimensional regime.

  \subsection{Overview}\label{sec::overview_framework}

  From a theoretical standpoint, the most challenging characteristic of a Hawkes process $\bm{N}$ is the inherent \emph{temporal dependence} in the intensity  \eqref{eqn::HP_intensity_general}.
  To be specific, the realization of $\bm{N}$ on any given time period $[t_1, t_2)$ depends on the realization of $\bm{N}$ on the previous time period $(-\infty,t_1)$.
  As a result, it is challenging to quantify the amount of information available in the observed realization of a Hawkes process on any period $[0,T]$.

  Most existing theoretical analyses of the Hawkes process rely on the cluster process representation proposed by \cite{hawkes1974}.
  The basic idea behind this representation is simple: for the linear Hawkes process \eqref{eqn::HP_intensity}, when the transfer functions  are non-negative, i.e., $\omega_{k,j}(\cdot) \geq 0$ for $1\leq j, k \leq p$, the process $\bm{N}$ can be represented as a sum of independent processes, or \emph{clusters}.
  Each cluster has intensity of the form \eqref{eqn::HP_intensity}, but with background intensity set to zero.
  Properties of the Hawkes process can then be investigated by studying the properties of independent clusters.
  See, among others, \cite{reynaud2007} and \cite{hansen2015} for recent applications of the cluster process representation. 

  Unfortunately, the cluster process representation is no longer available for the generalized Hawkes process \eqref{eqn::HP_intensity_general} in which $\phi_j(\cdot)$ may be non-linear and $\omega_{k,j}(\cdot)$ may be negative.
  To see this, note that a single cluster cannot model inhibition by itself, since its intensity is lower-bounded by zero.
  Moreover, independence across clusters implies that events in one cluster cannot affect the behavior of other clusters, which prohibits inhibition across clusters.
  Finally, the cluster process representation treats $\bm{N}$ as the \emph{summation} of clusters, which can only model an additive increase in the intensity, i.e., an intensity of the form \eqref{eqn::HP_intensity}.

  The lack of available techniques  to study the Hawkes process in the absence of linearity or non-negativity assumptions constitutes a significant gap between theory and applications of the Hawkes processes.
  As an example, networks of neurons are  known to have both excitatory and inhibitory connections \citep{van1996chaos, vogels2011inhibitory}.
  Similarly, it is unrealistic for neurons to have unbounded firing rates, which is possible for the linear Hawkes process.
  In fact, almost all applications of  the Hawkes process to neuronal spike train data use a non-linear link function  and avoid constraints on the signs of the transfer functions \citep{pillow2008,quinn2010,mishchenko2011,vidne2012,song2013}.
   To bridge the gap between the theory and emerging applications of the Hawkes process, we present a new approach to study theoretical properties of the Hawkes process without assuming that link functions are linear or that transfer functions are non-negative.

  The key idea of our new approach is to represent the generalized Hawkes process with intensity \eqref{eqn::HP_intensity_general} using the thinning process representation (see e.g., \citealt{ogata1981,bremaud1996}), which was first introduced by
  \cite{ogata1981} in order to simulate data from the Hawkes process.
  The thinning process representation has a clear advantage over the cluster process representation in that the former does not require the transfer functions to be non-negative or the link function to be linear.
  However, the thinning process representation has not been put into full use for the Hawkes process.
  In what follows, we will show that the thinning process representation can be used, in conjunction with a coupling result of \citet{dedecker2004},    to bound the temporal dependence of generalized Hawkes processes, without assuming linearity of ${\phi}_j(\cdot)$ or non-negativity of ${\omega}_{k,j}(\cdot)$.

  \subsection{Coupling Process Construction using Thinning}\label{sec::coupling_framework}

  To make the discussion more concrete, we consider the task of establishing a concentration inequality for 
  \begin{equation}\label{eqn::ybar}
  \bar{y}_{k,j} \equiv \frac{1}{T} \int_0^T \int_0^T f(t-t') \, \mathrm{d} N_k(t) \, \mathrm{d} N_j(t'), \quad 1\leq j, k\leq p,
  \end{equation}
  where $f(\cdot) $ is a known function with properties to be specified later.
  Quantities of  the form~\eqref{eqn::ybar} appear in many areas of  statistics,  such as regression analysis, cluster analysis, and principal components analysis.
  Concentration inequalities for~\eqref{eqn::ybar} thus provide the foundation for the theoretical analysis of these methods.

  Let
  \begin{equation}\label{eqn::y_kji}
  	{y}_{k,j,i} \equiv \frac{1}{ 2\epsilon } \int_{2\epsilon (i-1)}^{2\epsilon i} \int_0^T f(t-t') \, \mathrm{d} N_{k}(t)  \, \mathrm{d} N_j(t'),
  \end{equation}
  where $\epsilon$ is some small constant.
  For simplicity, assume that $T/(2\epsilon)$ is an integer.
  Then, $\bar{y}_{k,j}$ can be intuitively seen as the average of the sequence $\{{y}_{k,j,i} \}_{i=1}^{T/(2\epsilon)}$.
  Due to the nature of the Hawkes process, it is clear that the sequence $\{{y}_{k,j,i} \}_{i=1}^{T/(2\epsilon)}$ is inter-dependent, meaning that elements in the sequence depend on each other.
  As a result, standard concentration inequalities that require  independence do not apply to $\bar{y}_{k,j}$.
  Moreover, the Hawkes process is not a Markov process (outside of some special cases, e.g., a linear Hawkes process with exponential transfer functions).
  Thus, concentration inequalities for Markov processes do not apply to $\bar{y}_{k,j}$ either.
   Existing concentration inequalities for $|\bar{y}_{k,j} - \mathbb{E}\bar{y}_{k,j}|$ (see, e.g., Proposition~5 in \cite{reynaud2010}, Proposition~3 in \cite{hansen2015}) rely heavily on the cluster process representation \citep{hawkes1974,reynaud2007}, which, as discussed earlier, is not applicable due to the non-linearity of $\phi_j$ and the possibility of $\omega_{k,j}$ taking negative values.

  To develop a concentration inequality for $\bar{y}_{k,j}$, we bound the temporal dependence of the sequence $\{{y}_{k,j,i} \}_{i=1}^{T/(2\epsilon)}$ using the thinning process representation,  combined with the coupling result of \citet{dedecker2004}.

  First, we need to choose a measure of dependence for $\{{y}_{k,j,i} \}_{i=1}^{T/(2\epsilon)}$; see, e.g., the comprehensive survey by \cite{bradley2005}.
  We consider the $\tau$-dependence coefficient \citep{dedecker2004},
  \begin{equation}\label{eqn::tau_definition}
  \tau(\mathcal{M},X) \equiv \mathbb{E} \left[ \sup_{h} \left\{ \left| \int h(x) \mathbb{P}_{X\mid \mathcal{M}} (\mathrm{d} x) - \int h(x) \mathbb{P}_{X}( \mathrm{d} x) \right| \right\} \right],
  \end{equation}
  where the supremum is taken over all $1$-Lipschitz functions $h: \mathbb{R} \mapsto \mathbb{R}$, $X$ is a random variable, and $\mathcal{M}$ is a $\sigma$-field.
  Here, $\mathbb{P}_{X\mid \mathcal{M}}$ denotes the probability measure of $X$ conditioned on $\mathcal{M}$.
  We now introduce a coupling lemma from \cite{dedecker2004}.
  \begin{lemma}\label{lmm::tau_original}(Lemma 3 in \cite{dedecker2004})
  Let $X$ be an integrable random variable and $\mathcal{M}$   a $\sigma$-field defined on the same probability space.
  If the random variable $Y$  has the same distribution as $X$, and is independent of $\mathcal{M}$, then
  \begin{equation}\label{eqn::tau_coupling}
  \tau(\mathcal{M},X) \leq \mathbb{E} |X-Y|.
  \end{equation}
  \end{lemma}
  Lemma~\ref{lmm::tau_original} provides a practical approach for bounding the $\tau$-dependence coefficient: one can obtain an upper bound for $\tau(\mathcal{M},X)$ by constructing a coupling random variable $Y$, and evaluating $\mathbb{E} |X-Y|$.

  Equation~\ref{eqn::tau_definition} defines a measure of dependence between a random variable and a $\sigma$-field.
  For a sequence  $\{{y}_{k,j,i} \}_{i=1}^{T/(2\epsilon)}$, the temporal dependence is defined as (see, e.g., \citealp{merlevede2011})
  \begin{equation}\label{eqn::tau_sequence}
  \tau_{y}(l) \equiv \sup_{u\in\{1,2,\ldots\}} \tau\big( \mathcal{H}_{u}^{y},  y_{k,j,u+l} \big),
  \end{equation}
  for any positive integer $l$, where $\mathcal{H}_{u}^{y}$ is the $\sigma$-field determined by $\{{y}_{k,j,i} \}_{i=1}^{u}$, and the supremum in \eqref{eqn::tau_sequence} is taken over all positive integers.
  In words, for any time gap $l$, the temporal dependence of a sequence is defined  as the maximum dependence between any elements in the sequence and the history of the sequence $l$ steps ago.
  As a direct result of Lemma~\ref{lmm::tau_original}, we obtain the following coupling result for the temporal dependence \eqref{eqn::tau_sequence}:
  \begin{equation}\label{eqn::tau_coupling_sequence}
  \tau_{y}(l) \leq \sup_{u} \mathbb{E} \big|\tilde{y}_{k,j,u+l}^u  - y_{k,j,u+l}\big|,
  \end{equation}
    where $\{\tilde{y}_{k,j,i}^u \}_{i=1}^{T/(2\epsilon)}$ is a sequence satisfying the conditions of Lemma~\ref{lmm::tau_original}, i.e., $\{\tilde{y}_{k,j,i}^u \}_{i=1}^{T/(2\epsilon)}$ has the same distribution as $\{{y}_{k,j,i} \}_{i=1}^{T/(2\epsilon)}$, and is independent of $\{{y}_{k,j,i} \}_{i=1}^{u}$.

  Given the definition of temporal dependence \eqref{eqn::tau_sequence} and the coupling result \eqref{eqn::tau_coupling_sequence}, it remains to construct a coupling sequence  $\{\tilde{y}_{k,j,i}^u \}_{i=1}^{T/(2\epsilon)}$ and to bound the right-hand side of \eqref{eqn::tau_coupling_sequence}.
  Recalling that $y_{k,j,i}$ is of the form \eqref{eqn::y_kji}, the problem reduces to finding a coupling process $\widetilde{\bm{N}}^z$ and bounding the first- and second-order deviations between $\mathrm{d} \widetilde{\bm{N}}^z$ and $\mathrm{d} \bm{N}$, where $z \equiv 2\epsilon u$.
  In what follows, we will discuss the construction of  $\mathrm{d} \widetilde{\bm{N}}^z$ for a fixed $z$, and thus suppress the superscript $z$ for simplicity of notation.

  We use the thinning process representation of the Hawkes process to construct the coupling process $\mathrm{d} \widetilde{\bm{N}}$.
  To begin, we review the iterative construction strategy proposed by \cite{bremaud1996}.
  Let ${N}^{(0)}_j$, for $j=1,\ldots, p$,  be a homogeneous Poisson process on $\mathbb{R}^2$ with intensity $1$. For $n=1$,  we construct a $p$-variate process $\bm{N}^{(1)}$ as
  \begin{equation}\label{eqn::iterative_initial_main}
  \mathrm{d} N^{(1)}_j(t) = {N}^{(0)}_j\big( [0, \mu_j] \times \mathrm{d} t  \big)  \quad j=1,\ldots, p,
  \end{equation}
  where ${N}^{(0)}_j\big([0, \mu_j] \times \mathrm{d} t  \big)$ is the number of points for ${N}^{(0)}_j$ in the area $[0, \mu_j] \times [t, t+\mathrm{d} t)$.
  For $n \geq 2$, we construct $\bm{N}^{(n)}$ as
  \begin{equation}\label{eqn::iterative_construction_main}
  \begin{aligned}
  {\lambda}_j^{(n)}(t) & = \phi_j\big\{ {\mu}_j + \big( \bm{\omega}_{\cdot,j} * \mathrm{d} \bm{N}^{(n-1)} \big)(t) \big\} \\
  \mathrm{d} N^{(n)}_j(t) & = {N}^{(0)}_j\big( [0, \lambda_j^{(n)}(t)] \times \mathrm{d} t  \big), \quad j=1,\ldots, p.
  \end{aligned}
  \end{equation}
  \cite{bremaud1996} show that, under Assumption~\ref{asmp::spectralradius}, the sequence $\{ \bm{N}^{(n)} \}_{n=1}^{\infty}$ converges to the Hawkes process $\bm{N}$ with   intensity~\eqref{eqn::HP_intensity_general}.
  In a sense, \eqref{eqn::iterative_initial_main} and \eqref{eqn::iterative_construction_main} define the generalized Hawkes process for an arbitrary intensity function \eqref{eqn::HP_intensity_general}.

  In order to construct a coupling process, we modify the iterative construction strategy, \eqref{eqn::iterative_initial_main} and \eqref{eqn::iterative_construction_main}, as follows.
  Let $\widetilde{\bm{N}}^{(0)}$ also be a $p$-variate homogeneous Poisson process with each component process defined on $\mathbb{R}^2$ with intensity $1$, but independent of $\bm{N}^{(0)}$.
  For $j=1,\ldots, p,$ define the  process $\widetilde{\bm{N}}^{(1)}$ as
  \begin{equation}\label{eqn::iterative_initial_tilde_main}
  \mathrm{d} \widetilde{N}_j^{(1)}(t) =
  \begin{cases}
  \widetilde{N}^{(0)}_j\big( [0, \mu_j] \times \mathrm{d} t  \big) & t \leq z  \\
  {N}^{(0)}_j\big( [0, \mu_j] \times \mathrm{d} t  \big) & t > z
  \end{cases}.
  \end{equation}
  For $n \geq 2$, we construct $\widetilde{\bm{N}}^{(n)}$ as
  \begin{equation}\label{eqn::iterative_construction_tilde_main}
  \begin{aligned}
  \widetilde{\lambda}^{(n)}_j(t) & = \phi_j\big\{\mu_j + \big( \bm{\omega}_{\cdot,j} * \mathrm{d} \widetilde{\bm{N}}^{(n-1)} \big)(t)\big\} \\
  \mathrm{d} \widetilde{N}^{(n)}_j(t) & =
  \begin{cases}
  \mathrm{d} \widetilde{N}_j^{(0)}(  [0, \widetilde{\lambda}^{(n)}_j(t)]  \times \mathrm{d} t) & t \leq  z  \\
  \mathrm{d} N_j^{(0)}(  [0, \widetilde{\lambda}^{(n)}_j(t)]  \times \mathrm{d} t) & t >  z
  \end{cases}
  \quad j=1,\ldots, p.
  \end{aligned}
  \end{equation}

  The constructions of $\{\widetilde{\bm{N}}^{(n)}\}_{n=1}^{\infty}$ and $\{{\bm{N}}^{(n)}\}_{n=1}^{\infty}$ are almost identical, with the only difference being the use of $\widetilde{\bm{N}}^{(0)}$ or $\bm{N}^{(0)}$, which are identically distributed homogeneous Poisson processes.
  Thus, from \cite{bremaud1996}, we know that $\{\widetilde{\bm{N}}^{(n)}\}_{n=1}^{\infty}$ converges to a process $\widetilde{\bm{N}}$ with intensity~\eqref{eqn::HP_intensity_general}.
  As a result, the two processes $\widetilde{\bm{N}}$ and $\bm{N}$ are identically distributed.
  Thus, to apply Lemma~\ref{lmm::tau_original}, we only need to verify that $\widetilde{\bm{N}}$ is independent of $\mathcal{H}_z$.
  However, this is guaranteed by our construction, because the process ${\bm{N}}$ is determined by the homogeneous Poisson process $\bm{N}^{(0)}$ up to time $z$, which is independent of $\widetilde{\bm{N}}^{(0)}$ and also independent of $\mathrm{d} \bm{N}^{(0)}(t)$ for any $t > z$ due to properties of the homogeneous Poisson process.
  The next theorem shows that  the deviation between $\widetilde{\bm{N}}$ and ${\bm{N}}$ can be bounded.

  \begin{theorem}\label{thm::coupling}
  Let $\bm{N}$ be a Hawkes process with intensity of the form \eqref{eqn::HP_intensity_general} satisfying Assumption~\ref{asmp::spectralradius}.
  For any given $z >0$, there exists a point process $\widetilde{\bm{N}}$, called the coupling process of $\bm{N}$,  such that
  \begin{enumerate}
  \item[a)] $\widetilde{\bm{N}}$ has the same distribution as $ {\bm{N}}$;
  \item[b)] $\widetilde{\bm{N}}$ is independent of the history of $\bm{N}$ up to time $z$ (i.e., $\mathcal{H}_{z}$).
  \end{enumerate}
  Moreover, let $b$ be a constant and define a matrix $\bm{\eta}(b)=\big( \eta_{j,k}(b) \big)_{p\times p}$ with  $\eta_{j,k}(b) = \big[ \alpha_j \int_b^{\infty} |\omega_{k,j}(\Delta)| \mathrm{d} \Delta \big]$.
  Then, for any $u\geq 0$,
  \begin{align}
  \small
  \mathbb{E}\big| \mathrm{d} \widetilde{\bm{N}}(z+u) - \mathrm{d} {\bm{N}}(z+u) \big|/ \mathrm{d} u  & \preceq   2 v_1\left\{ \bm{\Omega}^{\lfloor u/b +1 \rfloor}  \bm{J} + \sum_{i=1}^{\lfloor u/b +1 \rfloor} \bm{\Omega}^{i-1} \bm{\eta}(b) \bm{J}\right\}, \label{eqn::bound_expectation}
  \end{align}
  \begin{equation}\label{eqn::bound_crosscovariance}
  \begin{aligned}
  \small
  & {\mathbb{E}\big| \mathrm{d} \widetilde{\bm{N}}(t')   \mathrm{d} \widetilde{\bm{N}}(z+u)  -  \mathrm{d} {\bm{N}}(t') \mathrm{d} {\bm{N}}(z+u) \big|}/\big({\mathrm{d} u \mathrm{d} t'} \big) \\
  \preceq  & 2 v_2 \left\{ \bm{\Omega}^{\lfloor u/b +2 \rfloor} + \sum_{i=1}^{\lfloor u/b +1 \rfloor} \bm{\Omega}^i \bm{\eta}(b) \right\}\bm{J}\bm{J}^{\T} + 2 v_1^2 \left\{  \bm{J}\bm{J}^{\T} \big(\bm{\Omega}^{\lfloor u/b +1 \rfloor}\big)^{\T}  + \sum_{i=1}^{\lfloor u/b +1 \rfloor} \bm{\eta}(b) \bm{J}\bm{J}^{\T} \big(\bm{\Omega}^{i}\big)^{\T}\right\},
  \end{aligned}
  \end{equation}
  where $\preceq$ denotes   element-wise inequality, and $v_1$ and $v_2$ are parameters that depend on $\bm{\Lambda}$, $\bm{V}$, and $\{\phi_j(\mu_j): j=1,\ldots,p\}$.
  \end{theorem}
  The proof of Theorem~\ref{thm::coupling} is given in Section~\ref{sec::coupling}.
  In both \eqref{eqn::bound_expectation} and \eqref{eqn::bound_crosscovariance}, the bounds are decomposed into two parts: terms involving $\bm{\eta}(b)$ (e.g., the last term in \eqref{eqn::bound_expectation} and the last two terms in \eqref{eqn::bound_crosscovariance}), and terms that do not involve $\bm{\eta}(b)$.
  By definition, $\bm{\eta}(b)$ characterizes the tail mass of $\omega_{k,j}, 1\leq j,k\leq p$, or,  in other words, the long-term \emph{direct} effect of an event.
  In contrast, $\bm{\Omega}^i$ captures the \emph{indirect} effect from the chain of events induced by an initial event.
  Intuitively, this decomposition reveals that the temporal dependence of a Hawkes process is jointly regulated by the long-term direct effects of every event and the indirect ripple effects descended from these events.

  \subsection{Main Results} \label{sec::highd}
  We now use the coupling process constructed in  Theorem~\ref{thm::coupling} to bound the weak dependence coefficient and  establish a concentration inequality for $\bar y_{k,j}$.
  To this end, we introduce two additional assumptions on the transfer functions $\omega_{k,j}(\cdot), 1\leq j,k \leq p$, which will allow us to  obtain element-wise bounds for the terms on the right-hand sides of both \eqref{eqn::bound_expectation} and \eqref{eqn::bound_crosscovariance}.
  To facilitate the analysis of the high-dimensional Hawkes process, we focus here on the setting where $p$ can grow.

  First of all, the bounds in both \eqref{eqn::bound_expectation} and \eqref{eqn::bound_crosscovariance} involve residual terms related to the tail behaviour of the transfer functions $\omega_{k,j}(\cdot)$. Naturally, the temporal dependence of a Hawkes process depends on the shape of the transfer functions, e.g., a process has stronger temporal dependence if effects of each event last longer.
  Therefore, we need an assumption on the tails of the transfer functions.
  \begin{assumption}\label{asmp::tail_highd}
  	There exists a constant $b_0$ such that, for $b\geq b_0$ and some $r>0$,
  	$$ \max_{j} \sum_{k=1}^p \int_{b}^{\infty} |\omega_{k,j}(\Delta)|\mathrm{d} \Delta \leq \constantTenP \exp\big(-\constantElevenP b^{r}\big).$$
  \end{assumption}
  This assumption prevents the long-term memory from dragging on indefinitely as $p$ increases.
  For instance, Assumption~\ref{asmp::tail_highd} is violated by a Hawkes process with transfer functions that equal zero except for $\omega_{k,k+1}(\Delta)= 0.5$ for $\Delta \in [k,k+1], k=1,\ldots,p-1$.
  We can see that the dependence between $\mathrm{d} N_{p-1}$ and $\mathrm{d} N_{p}$ grows as $p$ grows.
  The bounds on temporal dependence in \eqref{eqn::bound_expectation} and \eqref{eqn::bound_crosscovariance}, although still valid, is not meaningful in this setting as $p$ can grow.

  Secondly, in order to accommodate the high-dimensional regime, we further impose the following condition on the matrix $\bm{\Omega}$ defined in Assumption~\ref{asmp::spectralradius} of Section~\ref{sec::hawkes_rev}.
  \begin{assumption}\label{asmp::uniform}
  	For all $j=1,\ldots,p$, there exists a positive constant $\rho_{\Omega}<1$ such that  $\bm{\Omega}$ satisfies $\sum_{k=1}^p {\Omega}_{j,k} \leq \rho_{\Omega}$.
  \end{assumption}
  Applying Assumption~\ref{asmp::uniform} to \eqref{eqn::infinity_Lambda} for the linear Hawkes process (see Example~1) gives that, for each $j=1,\ldots,p$,
  \begin{equation}\label{eqn::infinity_Lambda_j}
  {\Lambda}_j =  \mu_j + \bm{\Omega}_{j,\cdot} \cdot \left[\sum_{i=1}^{\infty} \bm{\Omega}^{i-1} \bm{\mu}\right] =\mu_j + \bm{\Omega}_{j,\cdot} \cdot\bm{\Lambda} \leq \mu_j + \rho_{\Omega} \max_{k} (\Lambda_k).
  \end{equation}
  In a sense, Assumption~\ref{asmp::uniform} prevents the intensity from concentrating on a single process.
  Assumption~\ref{asmp::uniform} can be replaced by assumptions on the structure of $\bm{\Omega}$ if the magnitude of each entry of $\bm{\Omega}$ is upper bounded; for instance, it can be shown that Assumption~\ref{asmp::uniform} holds with high probability when the support of  $\bm{\Omega}$ corresponds to the adjacency matrix of a  sparse Erd\"{o}s-R\'{e}nyi graph or a stochastic block model with a suitable bound on $\max_{j,k} \Omega_{j,k}$ \citep{anandkumar2012}.

  With these additional assumptions, we arrive at the following bound for the temporal dependence coefficient.

  \begin{theorem}\label{thm::dependence}
  	Suppose that $\bm{N}$ is a Hawkes process with intensity~\eqref{eqn::HP_intensity_general}, which satisfies Assumptions~\ref{asmp::spectralradius}--\ref{asmp::bounded}.
  	Suppose also that the function $f(\cdot)$ has bounded support and $ \|f\|_{\infty}  \equiv \max_x |f(x)| \leq C_f$.
   For any positive integer $l$,  the $\tau$-dependence coefficient of $\{{y}_{j,k,i}\}_{i}$ introduced in \eqref{eqn::tau_sequence} satisfies
  		\begin{equation}\label{eqn::dependence_y}
  		\tau_y(l) \leq \constantSixteen \exp(-\constantSeventeen l^{r/(r+1)}),
  		\end{equation}
  		where $r$ is introduced in Assumption~\ref{asmp::tail_highd}, and $\constantSixteen$ and $\constantSeventeen$ are parameters that do not depend on $p$ and $T$.
  \end{theorem}

  In order to bound the deviation of  $\bar{y}_{j,k}$ from its mean, we need to introduce an additional assumption.
  \begin{assumption}\label{asmp::bounded}
  	Assume that one of the following two conditions holds.
  	\begin{itemize}
  		\item[a)] For all $j=1,\ldots,p$, the link functions $\phi_j(\cdot)$ in \eqref{eqn::HP_intensity_general} are upper-bounded by a positive constant $\phi_{\max}$.
  		\item[b)] In Assumption~\ref{asmp::tail_highd}, the constant $\constantTenP =0$.
  	\end{itemize}
  \end{assumption}
  Assumption~\ref{asmp::bounded} guarantees that the event count for the Hawkes process $N_j(A)$ has an exponential tail for any bounded interval $A$, as stated in the following lemma.
  \begin{lemma}\label{lmm::exponential_tail}
  	Suppose $\bm{N}$ is a Hawkes process with intensity~ \eqref{eqn::HP_intensity_general}, which satisfies Assumptions~\ref{asmp::spectralradius}~and~\ref{asmp::bounded}.
  	For any bounded interval $A$, it holds that, for $j=1,\ldots,p$, $P({{N}_j(A)}> n ) \leq \exp( 1- n/K)$ for some constant $K$.
  \end{lemma}
  The proof of Lemma~\ref{lmm::exponential_tail} is provided in Appendix~\ref{sec::proof_exp}.
  We hypothesize that Assumption~\ref{asmp::bounded} can be further relaxed, but we leave this for future work.

  Finally, we can establish the following concentration inequality for $\bar{y}_{j,k}$.
  \begin{theorem}\label{thm::concentration_hawkes}
  Suppose that $\bm{N}$ is a Hawkes process with intensity~\eqref{eqn::HP_intensity_general}, which satisfies Assumptions~\ref{asmp::spectralradius}--\ref{asmp::bounded}.
  Suppose also that the function $f(\cdot)$ has bounded support and $ \|f\|_{\infty}  \equiv \max_x |f(x)| \leq C_f$.
  Then, for $1 \leq j \leq k \leq p$,
  \begin{equation}\label{eqn::y_ci}
  \mathbb{P}\left(  \bigcap_{1 \leq j \leq k \leq p} \left[\left| \bar{y}_{k,j}  - \mathbb{E} \bar{y}_{k,j} \right| \geq  \constantFiveP T^{-(2r+1)/(5r+2)}  \right]\right) \leq  \constantSevenP p^2 T \exp( -\constantSixP T^{r/(5r+2)}),
  \end{equation}
  where $r$ is introduced in Assumption~\ref{asmp::tail_highd}, and $\constantFiveP, \constantSevenP$ and $ \constantSixP$ are parameters that do not depend on $p$ and $T$.
  \end{theorem}
  The proof of Theorem~\ref{thm::concentration_hawkes} is given in Section~\ref{sec::proofs_conc}, and is rather straightforward given Theorem~\ref{thm::dependence} and Lemma~\ref{lmm::exponential_tail}.
  Briefly, we verify that the conditions in \cite{merlevede2011} hold for the sequence  $\{{y}_{k,j,i} \}_{i=1}^{T/(2\epsilon)}$.
  We show that $y_{k,j,i}$ has an exponential tail of order $0.5$ using Lemma~\ref{lmm::exponential_tail}, and we show that the temporal dependence of the sequence  $\{{y}_{k,j,i} \}_{i=1}^{T/(2\epsilon)}$ decreases exponentially fast as in Theorem~\ref{thm::dependence}.
  Then, Equation \eqref{eqn::y_ci} is a direct result of Theorem~1 in \cite{merlevede2011}.

  \begin{remark}\label{rmk::costa}
  	In concurrent work, \cite{costa2018} have proposed an alternative approach for analyzing the one-dimensional Hawkes process with inhibition. They take $\phi(x) \equiv \max(0,x)$ and a transfer function with bounded support and possibly negative values.  Similar to our proposal, \cite{costa2018} employ the thinning process representation \citep{bremaud1996} to characterize the Hawkes process. However, the two proposals differ in the use of the thinning process representation.  \cite{costa2018} construct a linear Hawkes process $N^{+}$ that dominates  the original process $N$, which allows the use of the cluster process representation \citep{hawkes1974} and hence the existing theory on the linear Hawkes process. They further use $N^{+}$ to bound the renewal time of $N$, and establish limiting theorems using renewal techniques. In our proposal, we construct a coupling process $\widetilde{N}$ to bound the $\tau$-dependence coefficient of the generalized Hawkes process. This allows us to apply the theory of weakly dependent sequences (see, among others, Chapter~4 of \citealp{rio2017})  to the Hawkes process, and to obtain concentration inequalities for high-dimensional Hawkes processes with inhibition.
  \end{remark}

  \section{Application: Cross-Covariance Analysis of the Hawkes Process}\label{sec::cross-covariance}
  \subsection{Theoretical Guarantees}\label{sec::cross-covariance_theory}
  Cross-covariance analysis is widely used in the analysis of multivariate point process data.
  For instance, many authors have proposed and studied estimation procedures for the transfer functions $\bm{\omega}$ based on estimates of the cross-covariance $\bm{V}$ \citep{brillinger1976, krumin2010, bacry2014, etesami2016}.
  As another example, in neuroscience applications, it is common to cluster neurons based on the cross-covariances of their spike trains \citep{eldawlatly2009,feldt2010,muldoon2013,okun2015}.
  The empirical studies in \cite{okun2015} show that the $j$th neuron can be  viewed as a ``soloist"  or a ``chorister" based on estimates of ${\Lambda}_j^{-1} \sum_{k\neq j} {V}_{k,j}(0)$ for $j=1,\ldots,p$.

  However, estimators of the cross-covariance are not well-understood under practical assumptions. Existing theoretical studies often require  multiple realizations of the same process (see e.g., \citealp{bacry2014}) and non-negativity of the transfer functions. These assumptions do not always hold for real world point process data, such as financial data or neural spike train data.
  In this section, we demonstrate that Theorem~\ref{thm::concentration_hawkes} can be used to fill these gaps by providing a concentration inequality on smoothing estimators of the cross-covariance. As a concrete example, we consider the following smoothing estimator
  \begin{equation}\label{eqn::screening}
  \small
  \widehat{V}_{k,j}(\Delta) =  \begin{cases}
  (T h)^{-1} \iint_{[0,T]^2}  K \left( \frac{ (t' - t)+\Delta}{h} \right)\, \mathrm{d}  N_j(t') \mathrm{d}  N_{k}(t) -  T^{-1} N_j([0,T]) \frac{1}{{T}}N_{k}([0,T]) & j \neq k\\
  (T h)^{-1}\iint_{[0,T]^2 \backslash \{t=t' \}}   K \left( \frac{(t' - t)+\Delta}{h} \right)\, \mathrm{d}  N_k(t') \mathrm{d}  N_{k}(t) -  T^{-2} N^2_k([0,T]) & j = k\\
  \end{cases},
  \end{equation}
  where $K(\cdot)$ is a kernel function with bandwidth $h$ and  bounded support.

  Using Theorem~\ref{thm::concentration_hawkes}, we obtain the following concentration inequality for the smoothing estimator \eqref{eqn::screening}.
  \begin{corollary}
  	\label{cor::cross-covariance}
  	Suppose that $\bm{N}$ is a Hawkes process with intensity~\eqref{eqn::HP_intensity_general}, which satisfies Assumptions~\ref{asmp::spectralradius}--\ref{asmp::bounded}.
  	Further assume that the cross-covariances $\{V_{k,j},  1\leq j,k \leq p\}$ are $\theta_0$-Lipschitz functions.
  	Let $h=\constantEight T^{-(r+0.5)/(5r+2)}$ in \eqref{eqn::screening} for some constant $\constantEight$.
  	Then,
  	$$\mathbb{P}\left( \bigcap_{1 \leq j \leq k \leq p} \left[ \big\|\widehat{V}_{k,j}-V_{k,j} \big\|_{2,[-B,B]} \leq \constantEight T^{-\frac{r+0.5}{5r+2}} \right] \right) \geq 1-2\constantSevenP p^2 T^{\frac{6r+0.5}{5r+2}} \exp\left( - \constantSixP T^{\frac{r}{5r+2}} \right), $$
  	where $ \big\|\widehat{V}_{k,j}-V_{k,j} \big\|_{2,[-B,B]}^2\equiv \int_{-B}^{B} \big[\widehat{V}_{k,j}(\Delta)-V_{k,j}(\Delta)\big]^2 \mathrm{d} \Delta$ with $B$   a user-defined constant,  $\constantSevenP$ and $\constantSixP$ are constants introduced in Theorem~\ref{thm::concentration_hawkes}, and $\constantEight$ depends on $\theta_0$, $B$, and $\constantFiveP$ in Theorem~\ref{thm::concentration_hawkes}.
  \end{corollary}
  Corollary~\ref{cor::cross-covariance} provides a foundation for theoretical analysis of statistical procedures based on cross-covariances of the high-dimensional Hawkes process.
  It is a direct result of Theorem~\ref{thm::concentration_hawkes}. Its proof, given in Section~\ref{sec::proof_cross-covariance}, involves careful verification that the smoothing estimator \eqref{eqn::screening} satisfies the conditions in Theorem~\ref{thm::concentration_hawkes}.
  \subsection{Simulation Studies}\label{sec::simulation}

  In this section, we verify the theoretical result on estimators of the cross-covariance presented in Section~\ref{sec::cross-covariance_theory}.
  In all simulations, the intensity of the Hawkes process takes the form \eqref{eqn::HP_intensity_general} with $\phi_j(x)=\exp(x)/[1+\exp(x)]$, $\mu_j=1$, and $\omega_{k,j}(t) = a_{k,j}\gamma^{2}  t\exp(-\gamma t)$ for all $1\leq j,k\leq p$.  Here, the parameter  $\gamma$ controls the tail behavior of the transfer functions, and the parameter $a_{k,j}$ controls the magnitude of each transfer function. In what follows, we will provide details of the simulation setup and verify that the assumptions for  Corollary~\ref{cor::cross-covariance} are met.

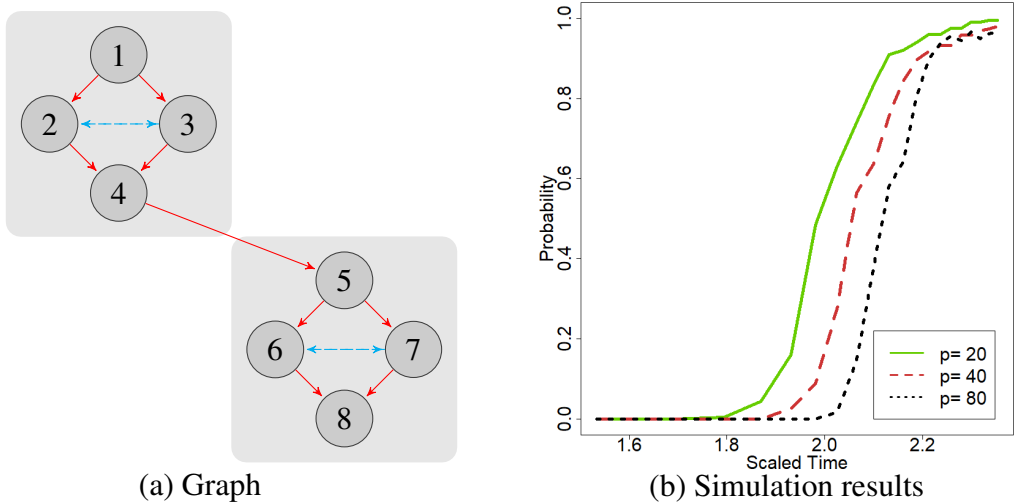
\begin{figure}
  	\centering
  \begin{tikzpicture}[node distance=1.3cm,>=stealth',bend angle=45,auto]
  \tikzstyle{inhub}=[circle,draw=black!75,fill=black!20,minimum size=6mm]
  \tikzstyle{outhub}=[circle,draw=black!75,fill=black!20,minimum size=6mm]
  \tikzstyle{cell}=[circle,draw=black!75,	fill=black!20,minimum size=6mm]

  \begin{scope}
  \node [inhub] (in1)                                    {1};
  \node [cell] (c1)  [below left of=in1] {2}
  edge [pre,color=red]                           (in1);
  \node [cell] (c2) [below right of=in1]                       {3}
  edge [pre,dashed,color=cyan]            (c1)
  edge [post,dashed,color=cyan]              (c1)
  edge [pre,color=red]                           (in1);
  \node [outhub] (out1) [below left of=c2]                      {4}
  edge [pre,color=red]                           (c1)
  edge [pre,color=red]                           (c2);
  \end{scope}

  \begin{scope}[xshift=3cm, yshift=-3cm]
  \node [inhub] (in1r)                                    {5}
  edge [pre,color=red] 							(out1);
  \node [cell] (c1r)  [below left of=in1r] {6}
  edge [pre,color=red]                           (in1r);
  \node [cell] (c2r) [below right of=in1r]                       {7}
  edge [pre,dashed,color=cyan]            (c1r)
  edge [post,dashed,color=cyan]              (c1r)
  edge [pre,color=red]                           (in1r);
  \node [outhub] (out1r) [below left of=c2r]                      {8}
  edge [pre,color=red]                           (c1r)
  edge [pre,color=red]                           (c2r);
  \node (title1)[below left of = out1r,xshift=-1cm] {(a) Graph };
  \end{scope}

  \begin{scope}[xshift=8.7cm, yshift=-2.5cm]
  \node (pic) {\includegraphics[width=6.5cm] {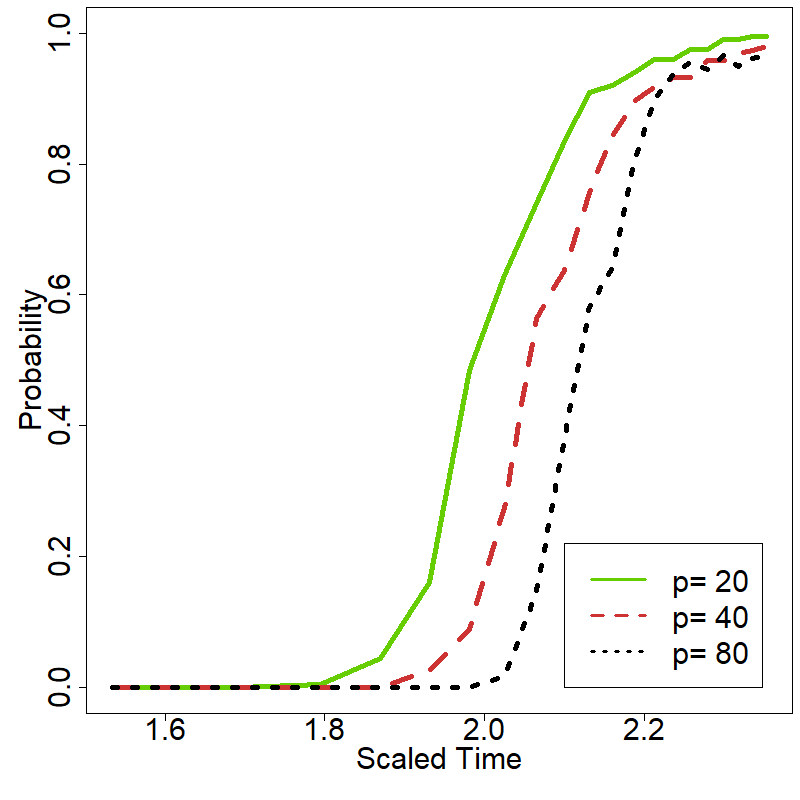}};
  \node [right of = title1,xshift=6.5cm] {(b) Simulation results };

  \end{scope}

  \begin{pgfonlayer}{background}
  \filldraw [line width=4mm,join=round,black!10]
  (in1.north  -| c2.east)  rectangle (out1.south  -| c1.west)
  (in1r.north  -| c2r.east)  rectangle (out1r.south  -| c1r.west);
  \end{pgfonlayer}
  \end{tikzpicture}
  	\caption{Simulation setup and results to verify the concentration inequality in Corollary~\ref{cor::cross-covariance}.  Panel (a) illustrates two blocks of the connectivity graph of the simulated Hawkes process. Each block contains four nodes, where only the first and the fourth nodes are connected to the nodes in the other blocks.  {Red solid arrows} indicate excitatory transfer functions, and  {blue dashed arrows} indicate inhibitory transfer functions. Self-inhibitory edges are omitted for ease of visualization. Panel (b) displays the empirical probability of the event $A$ in \eqref{eqn::event_A} over $200$ independent simulations. The $x$-axis displays the scaled time, $T^{1/7}$.  As expected from Corollary~\ref{cor::cross-covariance}, the empirical probability converges to unity exponentially fast as $T^{1/7}$ grows. }
  \label{fig::simulation}
  \end{figure}

  We generate networks of neurons as connected block-diagonal graphs shown in Figure~\ref{fig::simulation}(a). Each block consists of four nodes, among which Node 1 excites Nodes 2 and 3, and Nodes 2 and 3 are mutually inhibitory. Nodes 2 and 3 excite Node 4. The last node in each block excites the first node in the next block (i.e, $a_{4i+4,4i+5}>0$ for $i=0,1,2,...$). In addition, all nodes are self-inhibitory, i.e., $a_{j,j}<0$ for $j=1,\ldots p$.

  From the choice of $\omega_{k,j}$, we know that,  for any pair $(j,k)$, $\Omega_{j,k}=\|\omega_{k,j}\|_1 = a_{k,j}$ and  $\|\omega_{k,j}\|_{1,[b,\infty)}= a_{k,j} \exp(-b/\gamma)$. Thus, Assumption~\ref{asmp::tail_highd} is met with $r=1$.
  We set $a_{4i+2,4i+2}=a_{4i+3,4i+2}=a_{4i+3,4i+3}=a_{4i+2,4i+3}=a_{4i+4,4i+4}=-0.3$, and $a_{4i+2,4i+1}=a_{4i+3,4i+1}=a_{4i+4,4i+2}=a_{4i+4,4i+3}=0.3$ for the $i$th block, for $i=0,\ldots, p/4-1$. We set $a_{4i,4i+1}=0.45$ for $i=0,1,\ldots, p/4-2$.
  We further set $a_{1,1}=-0.9$ and $a_{4i+1,4i+1}=-0.45$ for $i=1,\ldots, p/4-1$. Noting that the link function $\phi_j(\cdot)$ is 1-Lipschitz, we can verify that $\bm{\Omega} $  satisfies Assumptions~\ref{asmp::spectralradius} and \ref{asmp::uniform}  with $\gamma_{\Omega}=0.9$ and $\rho_{\Omega}=0.9$. Finally, Assumption~\ref{asmp::bounded}(a) holds since $\phi_j(x) \leq \phi_{\max}=1$ for $j=1,\ldots, p$.

  We consider three scenarios with $p\in\{ 20, 40,  80\}$. In each scenario, we simulate a Hawkes process with $T$ ranging from $20$ to $400$, using the thinning process \citep{ogata1988}.  For each realization of the Hawkes process, we estimate the cross-covariance for all pairs of $1 \leq j,k \leq p$ using the estimator $\widehat{V}_{k,j}$ defined in  \eqref{eqn::screening} with  the Epanechnikov kernel and a bandwidth of $T^{-3/14}$ since $r=1$.  Since the analytical form of the true cross-covariance $\bm{V}$ \eqref{eqn::V} is unknown, we approximate the true cross-covariance with $\widetilde{\bm{V}}$, which is the average of $500$ smoothing estimators  on \emph{independent} realizations of the Hawkes process on $[0,200]$.
  By the law of large numbers, the estimator  $\widetilde{\bm{V}}$  is very close to the true value given the large number of independent samples, and we thus treat $\widetilde{\bm{V}}$ as the true cross-covariance in this numerical study.
  Define the event $A$ as
  \begin{equation}	\label{eqn::event_A}
  A\equiv\Big\{ \underset{{1\leq j,k \leq p}}{\medcap} \left[ \|\widehat{V}_{k,j}-\tilde{V}_{k,j}\|_{2,[-B,B]} \leq \constantEight T^{-3/14} \right]\Big\},
  \end{equation}
  where  $B=10$ and $\constantEight =0.32$.
  Corollary~\ref{cor::cross-covariance} states that the event \eqref{eqn::event_A} occurs with probability converging to unity as time $T$ increases.
  We will verify this claim by counting the proportion of cases that $A$ holds  in  $200$ independent simulations.

  Simulation results are shown in Figure~\ref{fig::simulation}(b).
  Note that $\constantEight$ does not have an analytic form in Corollary~\ref{cor::cross-covariance}.
  We evaluate the empirical probability of $A$ for a range of values of $\constantEight$, and  choose  $\constantEight=0.32$ to produce the curves in Figure~\ref{fig::simulation}.
  Other choices yield similar results.
  The $y$-axis displays the empirical probability of the event $A$ in \eqref{eqn::event_A} over $200$ simulations, and the $x$-axis displays the scaled time $T^{1/7}$. Corollary~\ref{cor::cross-covariance} claims that $\mathbb{P}(A)$ converges to unity exponentially fast as a function of $T^{1/7}$, as reflected in this plot. Furthermore, as $p$ increases, convergence slows down only slightly.

  \section{Proofs of Main Results}\label{sec::proofs_conc}

  In this section, we prove the main theoretical results from Sections~\ref{sec::conc} and \ref{sec::cross-covariance}.
  This section is organized as follows.
  In Section~\ref{sec::lemmas}, we list six technical lemmas,  and a theorem from \cite{merlevede2011} that are useful in the proofs.
  In Section~\ref{sec::coupling}, we prove Theorem~\ref{thm::coupling}, which guarantees the existence of a coupling process $\widetilde{\bm{N}}$ constructed in Section~\ref{sec::conc}, and bounds on the first- and second-order differences between $\mathrm{d}  \widetilde{\bm{N}}$ and $\mathrm{d} \bm{N}$.
  In Section~\ref{sec::weakdependence_y}, we prove Theorem~\ref{thm::dependence}, which bounds the $\tau$-dependence coefficient for the sequence $\{y_{k,j,i}\}_{i=1}^{T/\epsilon}$.
  In Section~\ref{sec::proofs_conc_main}, we apply the result in \cite{merlevede2011} to prove Theorem~\ref{thm::concentration_hawkes}, which establishes a concentration inequality for $\bar{y}_{j,k}$ in \eqref{eqn::ybar}.
  In Section~\ref{sec::proof_cross-covariance}, we prove the concentration inequality of cross-covariance estimators in  Corollary~\ref{cor::cross-covariance}.


  \subsection{Technical Lemmas}\label{sec::lemmas}
  The first two lemmas bound the expected deviation between the limiting processes of the two sequences constructed in Equations~\ref{eqn::iterative_initial_main}~and~\ref{eqn::iterative_construction_main}  and Equations~\ref{eqn::iterative_initial_tilde_main}~and~\ref{eqn::iterative_construction_tilde_main}, respectively.
   Note that, under Assumption~\ref{asmp::spectralradius}, the existence of the limiting process is shown by \cite{bremaud1996}.

  \begin{lemma}\label{lmm::bound_expectation}
  	Suppose that Assumption~\ref{asmp::spectralradius} holds for the intensity function \eqref{eqn::HP_intensity_general}.
  	Let $\bm{N}$ and $\widetilde{\bm{N}}$ be the limiting processes of the sequences  $\{{\bm{N}}^{(n)}\}_{n=1}^{\infty}$ and $\{\widetilde{\bm{N}}^{(n)}\}_{n=1}^{\infty}$, respectively.
  	Let $b$ be a constant and define a matrix $\bm{\eta}(b)=\big( \eta_{j,k}(b) \big)_{p\times p}$ with  $\eta_{j,k}(b) = \big[ \alpha_j \int_b^{\infty} |\omega_{k,j}(\Delta)| \mathrm{d} \Delta \big]$.
  	Then, for any $z>0$ and $u\geq 0$,
  	\begin{align*}
  	\small
  	\mathbb{E}\big| \mathrm{d} \widetilde{\bm{N}}(z+u) - \mathrm{d} {\bm{N}}(z+u) \big|/ \mathrm{d} u  & \preceq   2 v_1\left\{ \bm{\Omega}^{\lfloor u/b +1 \rfloor}  \bm{J} + \sum_{i=1}^{\lfloor u/b +1 \rfloor} \bm{\Omega}^{i-1} \bm{\eta}(b) \bm{J}\right\},
  	\end{align*}
  	where $\preceq$ denotes   element-wise inequality, and $v_1$ is a parameter that depends on $\bm{\Lambda}$ and $\{\phi_j(\mu_j): j=1,\ldots,p\}$.
  \end{lemma}

  \begin{lemma}\label{lmm::bound_crosscovariance}
  		Suppose that Assumption~\ref{asmp::spectralradius} holds for the intensity function \eqref{eqn::HP_intensity_general}.
  	Let $\bm{N}$ and $\widetilde{\bm{N}}$ be the limiting processes of the sequences  $\{{\bm{N}}^{(n)}\}_{n=1}^{\infty}$ and $\{\widetilde{\bm{N}}^{(n)}\}_{n=1}^{\infty}$, respectively.
  	Let $b$ be a constant and define a matrix $\bm{\eta}(b)=\big( \eta_{j,k}(b) \big)_{p\times p}$ with  $\eta_{j,k}(b) = \big[ \alpha_j \int_b^{\infty} |\omega_{k,j}(\Delta)| \mathrm{d} \Delta \big]$.
  	Then, for any $z>0$ and $u\geq 0$,
  	\begin{equation*}
  \begin{aligned}
  \small
  & {\mathbb{E}\big| \mathrm{d} \widetilde{\bm{N}}(t')   \mathrm{d} \widetilde{\bm{N}}(z+u)  -  \mathrm{d} {\bm{N}}(t') \mathrm{d} {\bm{N}}(z+u) \big|}/\big({\mathrm{d} u \mathrm{d} t'} \big) \\
  \preceq  & 2 v_2 \left\{ \bm{\Omega}^{\lfloor u/b +2 \rfloor} \bm{J}\bm{J}^{\T} + \sum_{i=1}^{\lfloor u/b +1 \rfloor} \bm{\Omega}^i \bm{\eta}(b) \bm{J}\bm{J}^{\T}\right\} + 2 v_1^2 \left\{  \bm{J}\bm{J}^{\T} \big(\bm{\Omega}^{\lfloor u/b +1 \rfloor}\big)^{\T}  + \sum_{i=1}^{\lfloor u/b +1 \rfloor} \bm{\eta}(b) \bm{J}\bm{J}^{\T} \big(\bm{\Omega}^{i}\big)^{\T}\right\},
  \end{aligned}
  \end{equation*}
  where $\preceq$ denotes   element-wise inequality,  $v_1$ is introduced in Lemma~\ref{lmm::bound_expectation}, and $v_2$ is a parameter that depends on $\bm{\Lambda}$, $\bm{V}$, and $\{\phi_j(\mu_j): j=1,\ldots,p\}$.
  \end{lemma}

  Proofs of Lemmas~\ref{lmm::bound_expectation}~and~\ref{lmm::bound_crosscovariance} are provided in Appendices~\ref{sec::proof_expectation}~and~~\ref{sec::proof_crosscovariance}, respectively.

  The next two lemmas provide bounds on every entry in the left-hand sides of the inequalities in Lemmas~\ref{lmm::bound_expectation}~and~\ref{lmm::bound_crosscovariance}.
  \begin{lemma}\label{lmm::elementwise_expectation}
  Suppose $\bm{N}$ is a Hawkes process with intensity~\eqref{eqn::HP_intensity_general}, which satisfies Assumptions~\ref{asmp::spectralradius}--\ref{asmp::uniform}.
  	For a given $z >0$, consider the process $\widetilde{\bm{N}}$ constructed in Theorem~\ref{thm::coupling}. Then, for $j=1,\ldots, p$,
  	$$\mathbb{E}\big|\mathrm{d} \widetilde{{N}}_j(z+u) - \mathrm{d} {{N}}_j(z+u) \big|/\mathrm{d} u \leq \constantTwelveP v\exp\big(-\constantThirteenP u^{r/(r+1)}\big),$$
  	where $r$ is defined in Assumption~\ref{asmp::tail_highd}, $v=\max(v_1,v_1^2, v_2)$ in Theorem~\ref{thm::coupling}, and $\constantTwelveP$ and $\constantThirteenP$ are parameters that do not depend on $p$ and $T$.
  \end{lemma}

  \begin{lemma}\label{lmm::elementwise_crosscovariance}
  Suppose $\bm{N}$ is a Hawkes process with intensity~\eqref{eqn::HP_intensity_general}, which satisfies Assumptions~\ref{asmp::spectralradius}--\ref{asmp::uniform}.
  For a given $z >0$, consider the process $\widetilde{\bm{N}}$ constructed in Theorem~\ref{thm::coupling}. Then, for $1\leq j,k\leq p$ and $t' > z+u$,
  $$  \mathbb{E}\big| \mathrm{d} \widetilde{{N}}_k(t')   \mathrm{d} \widetilde{{N}}_j(z+u)  -  \mathrm{d} {{N}}_k(t') \mathrm{d} {{N}}_j(z+u) \big|/(\mathrm{d} t' \mathrm{d} u)  \leq \constantFourteenP v \exp\big(-\constantFifteenP u^{r/(r+1)}\big), $$
  where $r$ is defined in Assumption~\ref{asmp::tail_highd}, $v=\max(v_1,v_1^2, v_2)$ in Theorem~\ref{thm::coupling}, and $\constantFourteenP$ and $\constantFifteenP$ are parameters that do not depend on $p$ and $T$.
  \end{lemma}
  We provide the proofs in Appendix~\ref{sec::proof_elementwise_expectation} and Appendix~\ref{sec::proof_elementwise_crosscovariance}, respectively.

  Lemma~\ref{lmm::poissontail} quantifies the tail behaviour of a Poisson random variable.
  We state it without proof.
  \begin{lemma}\label{lmm::poissontail}
  	Let $x$ be a Poisson random variable with mean $m$. Then for any $n>0$,
  	\begin{equation}
  	\mathbb{P}(x - m \geq n) \leq \exp\big(- m- [\log(n/m)-1] n \big).
  	\end{equation}
  \end{lemma}

  The next lemma characterizes the tail of the product of two random variables.
  \begin{lemma}\label{lmm::product}
  	Suppose that $Z_1$ and $Z_2$ satisfy
  	\begin{equation}
  	P(|Z_i|> n ) \leq \exp( 1- n/K_i), \quad i=1,2,
  	\end{equation}
  	for all $n \geq 0$.
  	Then, for any $n\geq 0$ and $K^*={K_1 K_2}(\log 2 + 1)$,
  	\begin{equation}
  	P(|Z_1 Z_2|> n ) \leq \exp\big( 1- (n/K^*)^{1/2}\big).
  	\end{equation}
  \end{lemma}
  The proof of Lemma~\ref{lmm::product} is provided in Appendix~\ref{sec::proof_product}.

  The following theorem from \cite{merlevede2011} will be used in the proof of Theorem~\ref{thm::concentration_hawkes}.

  \begin{theorem}\label{thm::thm_mpr}(Theorem~1 in \citet{merlevede2011}) Let $\{y_i\}_{i \in \mathbb{Z}}$ be
  	a sequence of  real valued random variables and let $v_y$ be defined as
  	\begin{equation}\label{eqn::v_y}
  	v_y = \sup_{i >0} \left\{ \mathbb{E}\big[ (y_{i}-\mathbb{E} y_{i})^2 \big] + 2\sum_{l \geq 1} \mathbb{E}\big[(y_{i}-\mathbb{E}y_{i})(y_{i+l}-\mathbb{E}y_{i+l})  \big] \right\}.
  	\end{equation}
  	Assume that
  	\begin{equation} \label{eqn::tau_thm}
  	\tau(y) \leq a \exp(-c x^{\gamma_1} ) \ {\rm for \ any \ } x \geq 1,
  	\end{equation}
  	and
  	\begin{equation}\label{eqn::tail_thm}
  	 \sup_{k > 0} \mathbb{P}(|y_k| >t) \leq \exp\big(1-(t/b)^{\gamma_2} \big).
  	\end{equation}
  	Further assume that $\gamma <1$ where $1/\gamma = 1/\gamma_1 + 1/\gamma_2$.
  	Then $v_y$ is finite and, for any $n\geq 4$, there exist positive constants $C_1$, $C_2$, $C_3$ and $C_4$ depending only on $a$, $b$, $c$, $\gamma$ and $\gamma_1$ such that, for any positive $x$,
  	\begin{equation}\label{eqn::concofy_thm}
  	\begin{aligned}
  	\mathbb{P}\left(\left| \frac{1}{n}\sum_{i=1}^{n} {y}_{i}  - \mathbb{E} {y}_{i} \right| \geq x \right) \leq &  n \exp\left( -\frac{ x^{\gamma}}{C_1} \right) +\exp\left(-\frac{x^2}{C_2(1+n v_y)} \right) \\
  	& +\exp\left[ -\frac{x^2 }{ C_3 n} \exp\left(\frac{x^{\gamma(1-\gamma)} }{C_4 \left[\log (x)\right]^{\gamma} } \right) \right].
  	\end{aligned}
  	\end{equation}
  \end{theorem}
  \subsection{Proof of Theorem~\ref{thm::coupling}}\label{sec::coupling}

  Without loss of generality, we assume that the link function $\phi_j(\cdot)$ is $1$-Lipschitz, $\alpha_j=1$ for all $j$. This can be achieved by setting $\tilde{\phi}_j(x)\equiv \phi_j(x/\alpha_j)$, $\tilde{\mu}_j = \mu_j \alpha_j$,  and $\tilde{\omega}_{k,j}(\Delta)=\alpha_j \omega_{k,j}(\Delta)$ for all $j,k$. Using this simplification, Assumption~\ref{asmp::spectralradius} means that ${\Omega}_{j,k} = \int_0^{\infty} |\omega_{k,j}(\Delta) |\mathrm{d} \Delta$ and $\Gamma_{\max}(\bm{\Omega})=\gamma_{\Omega}<1$.

  As outlined in Section~\ref{sec::coupling_framework}, the first part of the proof involves constructing two sequences of point processes that converge to $\bm{N}$ and its coupling process $\widetilde{\bm{N}}$, respectively.
  We have provided the construction in \eqref{eqn::iterative_construction_main} and  \eqref{eqn::iterative_construction_tilde_main}.
  Here we verify the two statements in Theorem~\ref{thm::coupling}: (a) $\widetilde{\bm{N}}$ has the same distribution as $ {\bm{N}}$, and  (b) $\widetilde{\bm{N}}$ is independent of the history of $\bm{N}$ up to time $z$.

  We notice that (a) in Theorem~\ref{thm::coupling} is a direct result of our construction.
  \cite{bremaud1996} show that, under Assumption~\ref{asmp::spectralradius}, the sequence $\big\{ \bm{N}^{(n)} \big\}_{n=1}^{\infty}$ converges to a process that has the same distribution as $\bm{N}$, i.e., the limiting process has intensity \eqref{eqn::HP_intensity_general}.
  In this proof, we will not distinguish between $\bm{N}$ and the limit of  $\big\{ \bm{N}^{(n)} \big\}_{n=1}^{\infty}$.
  We claim that the sequence  $\big\{ \widetilde{\bm{N}}^{(n)} \big\}_{n=1}^{\infty}$  also converges to a process with intensity \eqref{eqn::HP_intensity_general}.
  To see this, define a hybrid process $\widehat{{N}}_j^{(0)}(\mathrm{d} s \times \mathrm{d} t ) \equiv \indic_{[ t\leq z]} \widetilde{{N}}_j^{(0)} (\mathrm{d} s \times \mathrm{d} t )  + \indic_{[ t> z]} {{N}}_j^{(0)} (\mathrm{d} s \times \mathrm{d} t) $.
  The process $\widehat{{N}}_j^{(0)}$ is a homogeneous Poisson process on $\mathbb{R}^2$ with intensity $1$.
  This follows from the fact that for any Borel set $A \in \mathcal{B}(\mathbb{R}^2)$, $\widehat{{N}}_j^{(0)}(A)$ follows a Poisson distribution with expectation $m(A)$, where $m(A)$ is the Lebesgue measure of $A$.
  We can rewrite the construction of  $\big\{ \widetilde{\bm{N}}^{(n)} \big\}_{n=1}^{\infty}$ using $\widehat{\bm{N}}^{(0)}$ as
  \begin{equation}\label{eqn::iterative_initial_tilde2}
  \mathrm{d} \widetilde{N}_j^{(1)}(t) =
  \widehat{N}^{(0)}_j\big( [0, \mu_j] \times \mathrm{d} t  \big) \end{equation}
  and  for $n \geq 2$,
  \begin{equation}\label{eqn::construction_tilde2}
  \begin{aligned}
  \widetilde{\lambda}_j^{(n)}(t) & = \phi_j \left\{ \mu_j + \Big( \bm{\omega}_{\cdot,j} * \mathrm{d} \widetilde{\bm{N}}^{(n-1)} \Big)(t) \right\}\\
  \mathrm{d} \widetilde{N}^{(n)}_j(t) & =
  \mathrm{d} \widehat{N}_j^{(0)}\big(  [0, \widehat{\lambda}^{(n)}_j]  \times \mathrm{d} t\big).
  \end{aligned}
  \end{equation}
  The argument in \cite{bremaud1996} can thus be applied to Equations~\ref{eqn::iterative_initial_tilde2}~and~\ref{eqn::construction_tilde2}, which means that a limiting process $\widetilde{\bm{N}}$ exists with intensity of the form \eqref{eqn::HP_intensity_general}.

  To establish (b) in the statement of Theorem~\ref{thm::coupling}, we show that, for every $n$, $\widetilde{\bm{N}}^{(n)}$ is independent of $\mathcal{H}_z^{(n)}$.
  To see this, note that, by construction, $\widetilde{\bm{N}}^{(n-1)}$ is independent of the history of  $ {\bm{N}}^{(n-1)}$ before time $z$, denoted as $\mathcal{H}^{(n-1)}_z$.
  Since $\widetilde{\bm{\lambda}}^{(n)}$ is a function of $\widetilde{\bm{N}}^{(n-1)}$, we see that $\widetilde{\bm{\lambda}}^{(n)}$ is independent of $\mathcal{H}^{(n-1)}_z$.
  Thus, since $\widetilde{\bm{N}}^{(n)}$ is determined by $\widetilde{\bm{\lambda}}^{(n)}$ and $ \widetilde{\bm{N}}^{(0)}$, it is also independent of  $\mathcal{H}^{(n-1)}_z$.
  Hence, given that $\mathcal{H}^{(n)}_z$ is determined only by $\mathcal{H}^{(n-1)}_z$ and  $\bm{r}^{(n)}$,  $\widetilde{\bm{N}}^{(n)}$ is also independent of  $\mathcal{H}^{(n)}_z$.
  Therefore, the iterative construction preserves the independence between $\widetilde{\bm{N}}^{(n)}$ and  $\mathcal{H}^{(n)}_z$.
  As a result, $\widetilde{\bm{N}} \equiv \widetilde{\bm{N}}^{(\infty)} $ is independent of $\mathcal{H}^{(\infty)}_{z}\equiv \mathcal{H}_{z}$.

  So far, we have verified claims (a) and (b) of Theorem~\ref{thm::coupling}, i.e.,  that (a) there exist identically distributed ${\bm{N}}$ and $\widetilde{\bm{N}}$, and that (b) $\widetilde{\bm{N}}$ is independent of $\mathcal{H}_z$.

  The rest of the proof follows from Lemmas~\ref{lmm::bound_expectation}~and~\ref{lmm::bound_crosscovariance}, which lead to the inequalities in \eqref{eqn::bound_expectation}~and~\eqref{eqn::bound_crosscovariance}, respectively.  \QEDB

  \subsection{Proof of Theorem~\ref{thm::dependence}}\label{sec::weakdependence_y}

  Without loss of generality,  in this proof, we assume that $\text{supp}(f) \subset [-b_f, 0]$ for a positive constant $b_f$.
  It is straight-forward to generalize the results to $f$ with arbitrary bounded support.
  To see this, consider three scenarios for $[b_1,b_2] \equiv \text{supp}(f)$:
  \begin{enumerate}
  	\item $b_1<b_2 <0$. It is clear that $\text{supp}(f) \subset [b_f, 0] $ with $b_f \equiv b_1$.
  	\item $0<b_1 <b_2$. We can define $g(x)=f(-x)$. Then the proof applies to $\text{supp}(g)$.
  	\item $b_1<0<b_2$. We can write $f$ as $f=f^{+}+f^{-}$ where $f^{+}(x)=f(x)\indic_{[x>0]}$ and $f^{-}(x) = f(x) \indic_{[x\leq 0]}$.  Setting $g(x) = f^{+}(-x)$, we can see that the proof applies to $f^{-}$ and $g$, and thus to $f$.
  \end{enumerate}

  Recall that the series $\{{y}_{k,j,i}\}$ for $i=1,\ldots,T/(2\epsilon) $ is introduced in \eqref{eqn::y_kji} as
  \begin{equation*}
  {y}_{k,j,i} \equiv \frac{1}{ 2\epsilon } \int_{2\epsilon (i-1)}^{2\epsilon i} \int_0^T f(t-t') \mathrm{d} N_{k}(t) \mathrm{d} N_j(t').
  \end{equation*}
  Here, $\epsilon$ is the smallest number such that $\epsilon \geq \max\{b_f,b\}$ and $T/(2\epsilon)$ is an integer.

  We establish the bound \eqref{eqn::dependence_y} on the $\tau$-dependence coefficient of the sequence $\{y_{k,j,i}\}_{i=1}^{T/(2\epsilon)}$.
  For any $z$, we construct a sequence $\{\tilde{y}_{k,j,i}^z \}_{i=1}^{T/(2\epsilon)}$, with the same distribution as $\{y_{k,j,i} \}_{i=1}^{T/(2\epsilon)}$, but for all positive integers $l$, $\tilde{y}_{k,j,z+l}^z$ is independent of $\{y_{k,j,i} \}_{i=1}^z$.
  For simplicity, we suppress the superscript $z$ in the remainder of this proof.
  To this end, let  $\widetilde{\bm{N}}$ be the process in Theorem~\ref{thm::coupling} such that $\widetilde{\bm{N}}$ has the same distribution as $\bm{N}$  and $\widetilde{\bm{N}}$ is independent of $\mathcal{H}_{2\epsilon z}$.
  We define $\tilde{y}_{k,j,i}$ based on $\widetilde{\bm{N}}$ as
  \begin{equation}
  \tilde{y}_{k,j,i} \equiv \frac{1}{2\epsilon} \int_{2\epsilon (i-1)}^{2\epsilon i} \int_0^T f(t-t') \mathrm{d} \tilde{N}_{k}(t) \mathrm{d} \tilde{N}_j(t').
  \end{equation}
  From Theorem~\ref{thm::coupling}, we know that $\{\tilde{y}_{k,j,i} \}_{i=1}^{T/(2\epsilon)}$ has the same distribution as $\{y_{k,j,i} \}_{i=1}^{T/(2\epsilon)}$, and $\{\tilde{y}_{k,j,i}\}_{i=z+l}^{T/(2\epsilon)}$ is independent of $\{y_{k,j,i} \}_{i=1}^z$. We now bound the quantity $\mathbb{E} \big| \tilde{y}_{j,k,z+l} - {y}_{j,k,z+l}\big|$.

  For $l\geq (2\epsilon+b_f)/\epsilon$,
  \begin{equation*}
  \begin{aligned}
   \mathbb{E} \big| \tilde{y}_{j,k,z+l} - {y}_{j,k,z+l}\big|  = & \frac{1}{2\epsilon}\mathbb{E}  \left|\int_{2\epsilon(z+l-1)}^{2\epsilon(z+l)}\int_{t-b_f}^{t} f(t-t') \big[ \mathrm{d} \tilde{N}_k(t') \mathrm{d} \tilde{N}_j(t) - \mathrm{d} {N}_k(t') \mathrm{d} {N}_j(t) \big]\right| \\
  \leq &\frac{1}{2 \epsilon } \int_{2\epsilon(z+l-1)}^{2\epsilon(z+l)}\int_{t-b_f}^{t}  |f(t-t')| \mathbb{E}  \left| \mathrm{d} \tilde{N}_k(t') \mathrm{d} \tilde{N}_j(t) - \mathrm{d} {N}_k(t') \mathrm{d} {N}_j(t)\right| \\
  \leq &\frac{C_f}{2 \epsilon} \int_{2\epsilon(z+l-1)}^{2\epsilon(z+l)}\int_{t-b_f}^{t} \mathbb{E}  \left| \mathrm{d} \widetilde{N}_k(t') \mathrm{d} \widetilde{N}_j(t) - \mathrm{d} {N}_k(t') \mathrm{d} {N}_j(t)\right| \\
  \leq & \frac{C_f}{2 \epsilon} \int_{2\epsilon(z+l-1)}^{2\epsilon(z+l)}\int_{t-b_f}^{t}  \constantFourteenP p^{3/2}v_2\exp\big(-\constantFifteenP [ \min(t',t)-2\epsilon z]^{r/(r+1)}\big) \mathrm{d} t' \mathrm{d}t,
  \end{aligned}
  \end{equation*}
  where the second inequality follows from  $\|f\|_{\infty}\leq C_f<\infty$, and the last inequality follows from Lemma~\ref{lmm::elementwise_crosscovariance}.
  Given that $\min(t',t)-2\epsilon z\geq 2 \epsilon(z+l-1) -b_f -2\epsilon z = 2\epsilon (l-1) - b_f$ and $l\geq (2\epsilon+b_f)/\epsilon$, we have
  \begin{equation*}
  \mathbb{E} \big|\tilde{y}_{j,k,z+l}- {y}_{j,k,z+l}\big| \leq  \constantSixteenP  \exp(- \constantSeventeen l^{r/(r+1)}),
  \end{equation*}
  where $\constantSixteenP = C_f b_f p^{3/2} v a_3$ with $v_2$ introduced in Theorem~\ref{thm::coupling},  and $\constantSeventeen = \epsilon^{r/(r+1)} a_4$.

  For $l<(2\epsilon+b_f)/\epsilon$, we similarly have
  \begin{equation*}
  \mathbb{E} \big| \tilde{y}_{j,k,z+l} - {y}_{j,k,z+l}\big|   \leq C_f b_f \constantFourteenP p^{3/2} v.
  \end{equation*}
  Taking $\constantSixteen= \constantSixteenP \exp\big([2\epsilon+b_f]^{r/(r+1)} \big) $, for any positive integer $l$,  we arrive at
  \begin{equation*}
  \mathbb{E} \big|\tilde{y}_{j,k,z+l}- {y}_{j,k,z+l}\big| \leq  \constantSixteen \exp(- \constantSeventeen l).
  \end{equation*}
  Thus, using Eq.~\ref{eqn::tau_sequence} and \ref{eqn::tau_coupling_sequence},
  $$ \tau_y(l) \equiv \sup_{z} \tau\big( \mathcal{H}_{z}^{y},  y_{j,k, z+u} \big)  \leq  \constantSixteen \exp(- \constantSeventeen l^{r/(r+1)}), $$
  as required.  \QEDB

  \subsection{Proof of Theorem~\ref{thm::concentration_hawkes}}\label{sec::proofs_conc_main}
  %

  In what follows we use $C_1$, $C_2$, $C_3$, and $C_4$ to denote constants whose value might change from line to line.
  As in the proof of Theorem~\ref{thm::dependence}, we assume that $\text{supp}(f) \subset [-b_f, 0]$ for a positive constant $b_f$.

  We first verify that, for any $i$, ${y}_{k,j,i}$ in \eqref{eqn::y_kji} has an exponential tail of order $1/2$, i.e.,
  \begin{equation}\label{eqn::tail_ineq}
  \sup_{i >0} \mathbb{P}(|y_{k,j,i}| \geq x ) \leq \exp\left(1- \constantEightteen {x}^{1/2} \right),
  \end{equation}
  where $\constantEightteen$ is a constant.
  Since $\|f\|_{\infty} \leq C_f < \infty$, we know that
  \begin{equation}\label{eqn::y_naive_bound}
  {y}_{k,j,i} \leq \frac{C_f}{2\epsilon} N_j\big([2\epsilon(i-1), 2\epsilon i]  \big) N_k\big( [2\epsilon i -2 \epsilon -b_f, 2\epsilon i ) \big).
  \end{equation}
  Given that both $[2\epsilon(i-1), 2\epsilon i] $ and $ [2\epsilon i -2 \epsilon -b_f, 2\epsilon i )$ are  finite intervals, Lemma~\ref{lmm::exponential_tail} shows that  $N_j\big([2\epsilon(i-1), 2\epsilon i]  \big) $ and $ N_k\big( [2\epsilon i -2 \epsilon -b_f, 2\epsilon i ) \big)$ have exponential tails of order $1$.
  From Lemma~\ref{lmm::product}, we know that ${y}_{k,j,i}$ has an exponential tail of order $1/2$.

  From~\eqref{eqn::tail_ineq} and the conclusion of Theorem~\ref{thm::dependence}, we know that $\{y_{k,j,i}\}_{i=1}^{ T/(2\epsilon) }$ satisfies \eqref{eqn::tail_thm} and \eqref{eqn::tau_thm} in Theorem~\ref{thm::thm_mpr} with $\gamma_1=r/(r+1)$ and $\gamma_2=1/2$.
  Thus, applying Theorem~1 in \cite{merlevede2011} to $\{ {y}_{k,j,i}- \mathbb{E} y_{k,j,i}  \}_{i=1}^{T/(2\epsilon)}$ gives
  \begin{equation}\label{eqn::concofy}
  \begin{aligned}
  \mathbb{P}\left(\left| \sum_{i=1}^{ T/(2\epsilon) } {y}_{k,j,i}  - \frac{T}{2\epsilon} \mathbb{E} {y}_{k,j,i} \right| \geq T \epsilon_1 \right) \leq &  \frac{T}{2\epsilon} \exp\left( -\frac{ (\epsilon_1 T)^{r/(3r+1)}}{C_1} \right) +\exp\left(-\frac{\epsilon_1^2 T^2}{C_2(1+T v_y/ 2\epsilon)} \right) \\
  & +\exp\left[ -\frac{\epsilon_1^2 T^2}{ C_3 T/2\epsilon} \exp\left(\frac{(\epsilon_1 T)^{r(2r+1)/(3r+1)^2} }{C_4 \left[\log (\epsilon_1 T)\right]^{r/(3r+1)} } \right) \right],
  \end{aligned}
  \end{equation}
  where $\epsilon_1$ is to be specified later, and $v_y$ is a measure of the ``variance" of $y_{k,j,i}$, introduced in \eqref{eqn::v_y}.
  We can see that 
  \begin{equation}\label{eqn::v_y_bound}
  \small
  \begin{aligned}
  v_y = & \sup_{i >0} \left\{ \mathbb{E}\big[ (y_{k,j,i}-\mathbb{E} y_{k,j,i})^2 \big] + 2\sum_{l \geq 1} \mathbb{E}\big[ \mathbb{E}(y_{k,j,i+l}-\mathbb{E}y_{k,j,i+l} \mid y_{k,j,i})   (y_{k,j,i}-\mathbb{E}y_{k,j,i})\big] \right\}\\
  \leq & \sup_{i >0} \left\{ \mathbb{E}\big[ (y_{k,j,i}-\mathbb{E} y_{k,j,i})^2 \big] + 2\sum_{l \geq 1} \mathbb{E}\big[ \big|\mathbb{E}(y_{k,j,i+l}-\mathbb{E}y_{k,j,i+l} \mid y_{k,j,i})\big|   \big|( y_{k,j,i}-\mathbb{E}y_{k,j,i})\big|\big] \right\}\\
  \leq & \sup_{i >0} \left\{ \mathbb{E}\big[ (y_{k,j,i}-\mathbb{E} y_{k,j,i})^2 \big] + 2\sum_{l \geq 1} \mathbb{E}\big[ \constantSixteen \exp(-\constantSeventeen l)   \big( |y_{k,j,i}-\mathbb{E}y_{k,j,i}|\big)\big] \right\}\\
  \leq & \sup_{i >0} \left\{ \mathbb{E}\big[ (y_{k,j,i}-\mathbb{E} y_{k,j,i})^2 \big] + 2\sum_{l \geq 1} \constantSixteen \exp(-\constantSeventeen l)  \mathbb{E}\big( |y_{k,j,i}-\mathbb{E}y_{k,j,i}|\big) \right\}\\
  = & \sup_{i >0} \left\{ \mathbb{E}\big[ (y_{k,j,i}-\mathbb{E} y_{k,j,i})^2 \big] + 2  \mathbb{E}\big( |y_{k,j,i}-\mathbb{E}y_{k,j,i}|\big) \frac{\constantSixteen \exp(-\constantSeventeen)}{1-\exp(-\constantSeventeen)}  \right\},\\
  \end{aligned}
  \end{equation}
  where the second inequality follows from a property of $\tau$-dependence, which ensures that $\tau_{y}(l-m) \geq  |\mathbb{E}[y_{k,j,i+l}\mid y_{j,k,m}] - \mathbb{E}y_{k,j,i+l}|$ (see e.g., Equation 2.1 in \cite{merlevede2011}), and \eqref{eqn::dependence_y}. Furthermore, we know that both $\mathbb{E}\big[ (y_{k,j,i}-\mathbb{E} y_{k,j,i})^2 \big] $ and $\mathbb{E}|y_{k,j,i}-\mathbb{E}y_{k,j,i}|$ are finite since  ${y}_{k,j,i}$ has an exponential tail of order $1/2$. Therefore, $v_y$ is bounded.

  Letting $\epsilon_1 = \constantFiveP T^{-(2r+1)/(5r+2)}/2$ and noting that $\epsilon$ is fixed gives
  \begin{equation}\label{eqn::neweqinpfthm2}
  \mathbb{P}\left(\left| \bar{y}_{j,k}  - \mathbb{E} \bar{y}_{j,k} \right| \geq \constantFiveP T^{-(2r+1)/(5r+2)} \right) \leq  \constantSevenP T \exp( -\constantSixP T^{r/(5r+2)}),
  \end{equation}
  where the right-hand side of \eqref{eqn::neweqinpfthm2} dominates the terms on the right-hand side of \eqref{eqn::concofy}.
  Using a Bonferroni bound yields the result \eqref{eqn::y_ci} in Theorem~\ref{thm::concentration_hawkes}.
  \QEDB

  \subsection{Proof of  Corollary~\ref{cor::cross-covariance}}\label{sec::proof_cross-covariance}

  In the following, we only discuss the case $j\neq k$ for  $\widehat{V}_{k,j}$.
  The proof for $j = k$ follows from a similar argument and is omitted.

  Recall that the estimator $\widehat{V}_{k,j}$ takes the form (Equation~\ref{eqn::screening})
  \begin{equation}
  \widehat{V}_{k,j}(\Delta) =  \underbrace{\frac{1}{T h} \iint_{[0,T]^2}  K \left(  \frac{\{t' - t\}+\Delta}{h} \right)\, \mathrm{d}  N_j(t) \mathrm{d} N_{k}(t')}_{\text{I}/h}  -  \underbrace{\frac{1}{{T}} N_j(T) \frac{1}{{T}}N_{k}(T)}_{\text{II}}.
  \end{equation}

  For $\text{I}$, applying Theorem~\ref{thm::concentration_hawkes} with $f(x) = K \big(  [\Delta- \{t' - t\}]/h \big)$ gives
  \begin{equation}\label{eqn::concofy_II}
  \mathbb{P}\left(\left| \text{I}   - \mathbb{E} [\text{I} ] \right| \geq \constantFiveP  T^{-\frac{2r+1}{5r+2}} \right) \leq  \constantSevenP T \exp( -\constantSixP T^{\frac{r}{5r+2}}),
  \end{equation}
  where by \eqref{eqn::Lambda} and \eqref{eqn::V}
  \begin{equation}
  \begin{aligned}
  \mathbb{E} [ \text{I}  ] = &   \mathbb{E} \left[ \frac{1}{T}  \iint_{[0,T]^2}  K \left(  \frac{\{t' - t\}+\Delta}{h} \right)\, \mathrm{d} N_j(t) \mathrm{d} N_{k}(t') \right] \\
  = & \frac{1}{T}  \iint_{[0,T]^2}  K \left(  \frac{\{t' - t\}+\Delta}{h} \right) (V_{k,j}(t-t')+ \Lambda_j \Lambda_k) \mathrm{d} t \mathrm{d} t'.
  \end{aligned}
  \end{equation}
  Therefore,
  \begin{equation*}\label{eqn::bias_ini}
  \small
  \begin{aligned}
  & |\mathbb{E} [\text{I} ]- h [V_{k,j}( \Delta ) + \Lambda_j \Lambda_k ]| \\
  = & \left|   \frac{1}{T} \iint_{[0,T]^2}  K\left(\frac{t-t'+\Delta}{h} \right) \mathbb{E} [\mathrm{d} N_j(t') \mathrm{d} N_k(t) ]  - \frac{1}{T} \iint_{[0,T]^2}  K\left(\frac{t-t'+\Delta}{h} \right) [V_{k,j}(\Delta) + \Lambda_j \Lambda_k]  \mathrm{d} t \mathrm{d}t' \right| \\
  = & \left|   \frac{1}{T} \iint_{[0,T]^2}  K\left(\frac{t-t'+\Delta}{h} \right) \left\{\mathbb{E} [\mathrm{d} N_j(t') \mathrm{d} N_k(t) ] - \Lambda_j \Lambda_k \mathrm{d} t \mathrm{d} t'\right\}  - \right.\\
  & \left.   \frac{1}{T} \iint_{[0,T]^2}  K\left(\frac{t-t'+\Delta}{h} \right) V_{k,j}(\Delta)  \mathrm{d}t \mathrm{d}t' \right| \\
  = & \left|   \frac{1}{T} \iint_{[0,T]^2}  K\left(\frac{t-t'+\Delta}{h} \right) V_{k,j}(t'-t) \mathrm{d} t \mathrm{d} t'  - \frac{1}{T} \iint_{[0,T]^2}  K\left(\frac{t-t'+\Delta}{h} \right) V_{k,j}(\Delta)  \mathrm{d} t \mathrm{d} t' \right| \\
  = & \left|   \frac{1}{T} \iint_{[0,T]^2}  K\left(\frac{t-t'+\Delta}{h} \right) \big[V_{k,j}(t'-t) -V_{k,j}(\Delta)\big]  \mathrm{d} t \mathrm{d} t' \right|,
  \end{aligned}
  \end{equation*}
  where we use the definition of $\bm{V}$ in the third equality.
  Recalling that the kernel function $K(x/h)$ is defined on $[-h,h]$,  it follows that
  \begin{equation}\label{eqn::bias}
  \small
  \begin{aligned}
  & | \mathbb{E} [\text{I}] -  h [V_{k,j}(\Delta) + \Lambda_j \Lambda_k] | \\
  = & \left|   \frac{1}{T} \int_{0}^T \int_{\max(0, t+\Delta-h)}^{\min(T,t+\Delta+h)} K\left(\frac{t-t'+\Delta}{h} \right) \big[V_{k,j}(t'-t) -V_{k,j}(\Delta)\big]  \mathrm{d} t \mathrm{d} t' \right| \\
  \leq  & \left| \frac{1}{T} \int_{0}^T \int_{\max(0, t+\Delta-h)}^{\min(T,t+\Delta+h)} K\left(\frac{t-t'+\Delta}{h} \right) \theta_0  |t'-t-\Delta|  \mathrm{d} t \mathrm{d} t' \right| \\
  \leq  & \left|   \frac{1}{T} \int_{0}^T \int_{\max(0, t+\Delta-h)}^{\min(T,t+\Delta+h)} K\left(\frac{t-t'+\Delta}{h} \right) \theta_0 h  \mathrm{d} t \mathrm{d} t' \right| \\
  \leq & \left|   \frac{1}{T} \int_{0}^T  2\theta_0 h^2  \mathrm{d} t \right| \\
  = & 2 \theta_0  h^2,
  \end{aligned}
  \end{equation}
  where the first inequality follows from the fact that $V_{k,j}$ is a $\theta_0$-Lipschitz function, and the last inequality holds since the kernel function $K(t/h)$ integrates to $h$ on $\mathbb{R}$.

  Similarly, for each term in $\text{II}=N_j(T)N_k(T)/T^2$, an argument very similar to the proof of Theorem~\ref{thm::concentration_hawkes} gives, for $1 \leq j \leq p$, 
  \begin{equation}\label{eqn::concofN}
  \mathbb{P}\left(\left|  N_j(T)/T  - \Lambda_j \right| \geq \constantFiveP  T^{-\frac{2r+1}{5r+2}} \right) \leq  \constantSevenP T \exp( -\constantSixP T^{\frac{r}{5r+2}}).
  \end{equation}

  Combining \eqref{eqn::concofy_II}, \eqref{eqn::bias}, and \eqref{eqn::concofN}, we have, with probability at least  $1-2\constantSevenP T \exp( -\constantSixP T^{\frac{r}{5r+2}})$,
  \begin{equation*}
  \begin{aligned}
  \left| \widehat{V}_{k,j}(\Delta) -{V}_{k,j}(\Delta) \right| \leq & \left| h^{-1}{\rm I} -h^{-1}  \mathbb{E}[\text{I}] \right|+  \left| h^{-1} \mathbb{E}  [\text{I}]  - V_{j,k}(\Delta) + \Lambda_j \Lambda_k \right| + \\
  &  \left|\frac{1}{T^2}( N_j(T) -T \Lambda_j) N_{k}(T) \right| + \left|\Lambda_j \frac{1}{T} N_{k}(T) - \Lambda_j \Lambda_{k}  \right| \\
  \leq & \constantFiveP T^{-\frac{2r+1}{5r+2}} h^{-1} + 2\theta_0 h +  (\|\bm{\Lambda}\|_{\infty}+\constantFiveP T^{-\frac{2r+1}{5r+2}}) \constantFiveP T^{-\frac{2r+1}{5r+2}} + \|\bm{\Lambda}\|_{\infty} \constantFiveP T^{-\frac{2r+1}{5r+2}}.
  \end{aligned}
  \end{equation*}
  Thus, letting $h=\constantEight T^{-\frac{r+0.5}{5r+2}}$, we can see that, for some constant $\constantFivePrime$, 
  \begin{equation}
  \left| \widehat{V}_{k,j}(\Delta) -{V}_{k,j}(\Delta ) \right| \leq \constantFivePrime T^{-\frac{r+0.5}{5r+2}}.
  \end{equation}
  Lastly, we need a uniform bound on $\widehat{V}_{k,j}-V_{k,j}$ on the region $[-B,B]$. We first note that the probability statement \eqref{eqn::concofN} holds for a grid of $\lceil T^{\frac{r+0.5}{5r+2}}\rceil$ points on $[-B,B]$, denoted as $\{ \Delta_i \}_{i=1}^{\lceil T^{\frac{r+0.5}{5r+2}}\rceil}$.
  The gap between adjacent points on this grid is upper-bounded by $2B T^{\frac{r+0.5}{5r+2}}$.
  Furthermore, for any $\Delta \in [-B,B]$, we can find a point $\Delta_i$ on the grid  such that $| \Delta - \Delta_i| \leq 2B/ \lceil T^{\frac{r+0.5}{5r+2}} \rceil \leq 2B T^{-\frac{r+0.5}{5r+2}}$.
  From basic algebra,   for all $\Delta \in [-B,B]$, 
  \begin{equation*}
  \begin{aligned}
  \left| \widehat{V}_{k,j}(\Delta)-V_{k,j}(\Delta) \right| = & \left| \widehat{V}_{k,j}(\Delta)-\widehat{V}_{k,j}(\Delta_k)+\widehat{V}_{k,j}(\Delta_k)-{V}_{k,j}(\Delta_k) +{V}_{k,j}(\Delta_k)- V_{k,j}(\Delta) \right|\\
  \leq &  2B T^{-\frac{r+0.5}{5r+2}} +  \constantFivePrime T^{-\frac{r+0.5}{5r+2}} + 2\theta_0 BT^{-\frac{r+0.5}{5r+2}}  \\
  \leq & \constantFivePP T^{-\frac{r+0.5}{5r+2}},
  \end{aligned}
  \end{equation*}
  where $\constantFivePP = 2B+\constantFivePrime +2\theta_0 B$.

  Now, taking a union bound, with probability at least $1-2\constantSevenP p^2 T^{\frac{6r+0.5}{5r+2}} \exp\left( - \constantSixP T^{\frac{r}{5r+2}} \right)$, it holds for all $j,k$ that, for some constant $\constantEight$,
  $\big\|\widehat{V}_{k,j}-V_{k,j}\big\|_{2,[-B,B]} \leq \constantEight T^{-\frac{r+0.5}{5r+2}}.$   \QEDB

  \section{Discussion}\label{sec::discussion}

  The proposed approach in Section~\ref{sec::conc} generalizes existing theoretical tools for the Hawkes process by lifting the strict assumptions that allow only mutually-exciting relationships and linear link functions \citep{hawkes1974}.
  However, more challenges remain in the analysis of the Hawkes process with inhibitory relationships.
  For instance, the assumption for stability (Assumption~\ref{asmp::spectralradius}) introduced by \cite{bremaud1996} puts a strong restriction on the matrix $\bm{\Omega}$.
  Here, each entry of $\bm{\Omega}$ is the $\ell_1$-norm of the corresponding transfer function, which neglects its sign.
  This assumption can be too restrictive in the presence of inhibitory relationships.
  To see this, consider a bivariate Hawkes process with a self-regulatory function $\omega_{1,1}(\Delta) = \omega_{2,2}(\Delta) = -2a \bm{1}_{[\Delta < 1]}$, $\omega_{1,2}(\Delta)=\omega_{2,1}(\Delta)= a \bm{1}_{[\Delta < 1]}$, any positive $1-$Lipschitz link function, and a non-zero spontaneous rate $\mu_1=\mu_2=a$.
  For any $a\geq 1/3$, it is clear that this specification yields a stable process despite violating Assumption~\ref{asmp::spectralradius}. It is natural to hypothesize that the requirement for stability depends on the sign of the transfer functions and the graphical structure.  Relaxing this assumption could be a fruitful direction of future research.

\bibliographystyle{chicago}
\bibliography{paper-ref}

\newpage
\appendix

\section{Proofs of Technical Lemmas}\label{sec::proof_lemmas}

\subsection{Proof of Lemma~\ref{lmm::bound_expectation}}\label{sec::proof_expectation}

Recall from the statement of Lemma~\ref{lmm::bound_expectation} that we define $\bm{ \eta}(b)\equiv (\eta_{j,k})_{p \times p}$ where  $\eta_{k,j}(b) =\int_{b}^{\infty} |\omega_{k,j}(\Delta)| \mathrm{d} \Delta$, for $j,k \in \{1,\ldots, p\}$.
Since $b$ is a constant in this proof, we will denote $\bm{ \eta}(b)$ as $\bm{ \eta}$ for ease of notation.
We claim that, for  $u > (n-1)b$, $n=1,2,\ldots,$
\begin{equation}\label{eqn::tau_N_hypothesis}
\mathbb{E}\big|\mathrm{d} \widetilde{\bm{N}}(z+u) - \mathrm{d} {\bm{N}}(z+u) \big|/\mathrm{d} u \preceq   2 v_1 \left(\bm{\Omega}^n \bm{J} +  \sum_{i=1}^n \bm{\Omega}^{i-1}  \bm{\eta}\bm{J}\right),
\end{equation}
where $\preceq$ is the element-wise inequality, $\bm{J}$ is a $p$-vector of ones, and $v_1$ is a constant such that $v_1 = \max_j \max( \phi_j(\mu_j) , \Lambda_j)$.

From \cite{bremaud1996}, we know that the limits of $\{{\bm{N}}^{(n)}\}_{n=1}^{\infty}$ and $\{\widetilde{\bm{N}}^{(n)}\}_{n=1}^{\infty}$ exist under Assumption~\ref{asmp::spectralradius}.
Therefore, for $j=1,\ldots, p$, the limiting process  $\bm{N}$ satisfies, for $j=1,\ldots, p$,
\begin{equation} \label{eqn::N_infinity}
\begin{aligned}
{\lambda}_j(t) & = \phi_j \big\{  \mu_j + \big( \bm{\omega}_{\cdot,j} * \mathrm{d} {\bm{N}} \big)(t) \big\} \\
\mathrm{d} N_j(t) & =  \mathrm{d} {N}_j^{(0)}\big( [0, {\lambda}_j(t)] \times \mathrm{d} t\big).
\end{aligned}
\end{equation}
and $\widetilde{\bm{N}}$ satisfies
\begin{equation} \label{eqn::Ntilde_infinity}
\begin{aligned}
\widetilde{\lambda}_j (t) & = \phi_j \big\{\mu_j + \big( \bm{\omega}_{\cdot,j} * \mathrm{d} \widetilde{\bm{N}} \big)(t) \big\} \\
\mathrm{d} \widetilde{N}_j(t) & =   \indic_{[ t\leq z]}  \widetilde{N}_j^{(0)}\big( [0, \widetilde{\lambda}_j(t)] \times \mathrm{d} t\big) +  \indic_{[ t > z]} N_j^{(0)}\big([0, \widetilde{\lambda}_j(t)] \times \mathrm{d} t\big).
\end{aligned}
\end{equation}

Using Equations~\ref{eqn::N_infinity}~and~\ref{eqn::Ntilde_infinity}, we can then prove \eqref{eqn::tau_N_hypothesis} by induction.
First note that, for any $u \in \mathbb{R}$,
\begin{equation}\label{eqn::bound_N_crude}
\small
\mathbb{E}\big|\mathrm{d} N_j(u+z)- \mathrm{d} \widetilde{N}_j(u+z) \big|/\mathrm{d}u \leq  \mathbb{E}\big|\mathrm{d} N_j(u+z) \big|/\mathrm{d} u +  \mathbb{E}\big| \mathrm{d} \widetilde{N}_j(u+z) \big|/\mathrm{d} u = 2 \Lambda_j \leq 2 v_1,
\end{equation}
where $\Lambda_j$ is the marginal intensity of $\mathrm{d} N_j$ defined in \eqref{eqn::Lambda}.
Hence, jointly for $j=1,\ldots,p$, $\mathbb{E}\big|\mathrm{d} \bm{N}(u+z)- \mathrm{d} \widetilde{\bm{N}}(u+z) \big|/\mathrm{d} u \preceq 2 v_1\bm{J}$ for any $u \in \mathbb{R}$.
We will then establish the bound for $u > (m-1)b$ for $m \in \{1,2,\ldots\}$ and $b$ introduced in Theorem~\ref{thm::coupling}.

For $m=1$, i.e., when $ u > 0$, we have
\begin{equation*}
\begin{aligned}
& \mathbb{E}\big|\mathrm{d} {N}_j(u+z)- \mathrm{d} \widetilde{{N}}_j(u+z) \big|/\mathrm{d} u \\
= & \mathbb{E}\big|N_j^{(0)}([0, \widetilde{\lambda}_j(u+z)] \times \mathrm{d} t)  -
N_j^{(0)}([0, {\lambda}_j(u+z)] \times \mathrm{d} t) \big| / \mathrm{d}u \\
= &  \mathbb{E}\big| {\lambda}_j(u+z) - \widetilde{\lambda}_j(u+z) \big| \\
= &  \mathbb{E}\Big| \phi_j\Big\{\mu_j+ \sum_{l=1}^p \big({\omega}_{l,j} * \mathrm{d} {N}_{l}\big) (u+z) \Big\} - \phi_j\Big\{\mu_j+ \sum_{l=1}^p \big({\omega}_{l,j} * \mathrm{d} \widetilde{N}_{l}\big) (u+z) \Big\} \Big|,
\end{aligned}
\end{equation*}
where the second equality follows since the expected differences between event counts is the expected differences between areas.
Recalling that we assume $\phi_j(\cdot)$ to be $1$-Lipschitz, we have
\begin{equation*}
\begin{aligned}
& \mathbb{E}\Big| \phi_j\Big\{\mu_j+ \sum_{l=1}^p \big({\omega}_{l,j} * \mathrm{d} {N}_{l}\big) (u+z) \Big\} - \phi_j\Big\{ \mu_j+ \sum_{l=1}^p \big({\omega}_{l,j} * \mathrm{d} \widetilde{N}_{l}\big) (u+z) \Big\} \Big| \\
\leq & \mathbb{E}\left|   \sum_{l=1}^p \big({\omega}_{l,j} * \mathrm{d} {N}_{l}\big) (u+z)-\sum_{l=1}^p \big({\omega}_{l,j} * \mathrm{d} \widetilde{N}_{l}\big) (u+z)  \right|\\
= &  \mathbb{E}\left|  \sum_{l=1}^p \int_0^{\infty} \omega_{l,j}(\Delta) \big[ \mathrm{d} N_l(u+z-\Delta) - \mathrm{d} \widetilde{N}_l(u+z-\Delta) \big] \right| \\
\leq &  \sum_{l=1}^p \int_0^{\infty} |\omega_{l,j}(\Delta)|\,  \mathbb{E}\big| \mathrm{d} N_l(u+z-\Delta) - \mathrm{d} \widetilde{N}_l(u+z-\Delta) \big|.
\end{aligned}
\end{equation*}
For each $l$, we note that
\begin{equation*}
\begin{aligned}
& \int_0^{\infty} |\omega_{l,j}(\Delta)| \, \mathbb{E}\big| \mathrm{d} N_l(u+z-\Delta) - \mathrm{d} \widetilde{N}_l(u+z-\Delta) \big| \\
= &  \left\{ \int_0^{b} +\int_{b}^{\infty}\right\} |\omega_{l,j}(\Delta)| \,  \mathbb{E}\big| \mathrm{d} N_l(u+z-\Delta) - \mathrm{d} \widetilde{N}_l(u+z-\Delta) \big|\\
\leq & \int_0^{b} |\omega_{l,j}(\Delta)| \mathrm{d} \Delta  \left\{ \max_{t' \in [u-b,u]}\mathbb{E}\big| \mathrm{d} N_l(t') - \mathrm{d} \widetilde{N}_l(t') \big|/\mathrm{d} t' \right\} +\\
&  \int_{b}^{\infty} |\omega_{l,j}(\Delta)| \mathrm{d} \Delta  \left\{ \max_{t' \in [-\infty, u-b]}\mathbb{E}\big| \mathrm{d} N_l(t') - \mathrm{d} \widetilde{N}_l(t') \big|/\mathrm{d} t'  \right\}\\
\leq & \int_0^{b} |\omega_{l,j}(\Delta)| \mathrm{d} \Delta  \left\{ \max_{t' \geq  -b} \mathbb{E}\big| \mathrm{d} N_l(t') - \mathrm{d} \widetilde{N}_l(t') \big|/\mathrm{d} t'  \right\} +\\
& \int_{b}^{\infty} |\omega_{l,j}(\Delta)| \mathrm{d} \Delta  \left\{ \max_{t' \in \mathbb{R} }\mathbb{E}\big| \mathrm{d} N_l(t') - \mathrm{d} \widetilde{N}_l(t') \big| /\mathrm{d} t'  \right\}\\
\leq &  \int_0^{b} |\omega_{l,j}(\Delta)| \mathrm{d} \Delta  \big(\Lambda_l + \Lambda_l\big)  +  \int_{b}^{\infty} |\omega_{l,j}(\Delta)| \mathrm{d} \Delta   \big(\Lambda_l + \Lambda_l\big) \\
\leq &    2\Omega_{j,l} v_1 + 2 \eta_{j,l} v_1,
\end{aligned}
\end{equation*}
where the second-to-last inequality follows from \eqref{eqn::bound_N_crude}, and the last inequality follows from the definition of $v$ and the tail property of $\omega_{j,l}$.

Combining the above inequalities, we see that, for $m=1$,
\begin{equation*}
\mathbb{E}\big|\mathrm{d} {N}_j(u+z)- \mathrm{d} \widetilde{{N}}_j(u+z) \big|/\mathrm{d} u \leq  2 v_1 \big( \bm{\Omega}_{j,\cdot}^{\T} \bm{J}+ \bm{\eta}_{j,\cdot} \bm{J}\big),
\end{equation*}
Thus, jointly for all $j$ and for $ u >0$,
\begin{equation*}
\mathbb{E}\big|\mathrm{d} \bm{N}(u+z)- \mathrm{d} \widetilde{\bm{N}}(u+z) \big|/ \mathrm{d} u \preceq  2v_1 \big( \bm{\Omega} \bm{J} +\bm{\eta}\bm{J}\big),
\end{equation*}
i.e., \eqref{eqn::tau_N_hypothesis} holds for $m=1$.

Now assume that \eqref{eqn::tau_N_hypothesis} holds for $m=n-1$, i.e., when $ u > (n-2)b$,
\begin{equation*}
\mathbb{E}\big|\mathrm{d} \bm{N}(u+z)- \mathrm{d} \widetilde{\bm{N}}(u+z) \big|/ \mathrm{d} u \preceq  2 v_1 \left( \bm{\Omega}^{n-1} \bm{J}+  \sum_{i=1}^{n-1} \bm{\Omega}^{i-1} \bm{\eta}\bm{J}\right).
\end{equation*}

Then, for $u >(n-1)b$, we have
\begin{equation*}
\begin{aligned}
& \mathbb{E}\big|\mathrm{d}{N}_j(u+z)- \mathrm{d} \widetilde{{N}}_j(u+z) \big|/ \mathrm{d} u \\
\leq & \sum_{l=1}^p \int_0^{b} |\omega_{l,j}(\Delta)| \mathrm{d} \Delta  \left\{ \max_{t' \geq  (n-2)b} \mathbb{E}\big| \mathrm{d} N_l(t') - \mathrm{d} \widetilde{N}_l(t') \big|/\mathrm{d} t'  \right\} +  \\
& \sum_{l=1}^p \int_{b}^{\infty} |\omega_{l,j}(\Delta)| \mathrm{d} \Delta  \left\{ \max_{t' \in \mathbb{R} }\mathbb{E}\big| \mathrm{d} N_l(t') - \mathrm{d} \widetilde{N}_l(t') \big| /\mathrm{d} t'  \right\}\\
\leq & \sum_{l=1}^p \int_0^{b} |\omega_{l,j}(\Delta)| \mathrm{d} \Delta   \left[ 2 v_1 \big(\bm{\Omega}^{n-1}\big)_{l,\cdot} \bm{J} +2v_1   \sum_{i=1}^{n-1} \big(\bm{\Omega}^{i-1}\big)_{l,\cdot}  \bm{\eta}  \bm{J} \right]  +  \sum_{l=1}^p \int_{b}^{\infty} |\omega_{l,j}(\Delta)| \mathrm{d} \Delta  2v_1 \\
\leq & \sum_{l=1}^p \Omega_{j,l} \left[ 2 v_1 \big(\bm{\Omega}^{n-1}\big)_{l,\cdot} \bm{J} +2v_1   \sum_{i=1}^{n-1} \big(\bm{\Omega}^{i-1}\big)_{l,\cdot} \bm{\eta} \bm{J} \right] + 2v_1 \bm{\eta}_{j,\cdot} \bm{J}\\
= & 2v_1 \bm{\Omega}_{j,\cdot} \bm{\Omega}^{n-1} \bm{J} +2v_1 \sum_{i=1}^{n-1} \big(\bm{\Omega}^{i}\big)_{l,\cdot} \bm{\eta} \bm{J} + 2v_1 \bm{\eta}_{j,\cdot} \bm{J},
\end{aligned}
\end{equation*}
where the inequalities follow from a similar argument to the case of $m=1$.
Thus,  for $m=n$, i.e., $ u > (n-1)b$,
\begin{equation*}
\mathbb{E}\big|\mathrm{d} \bm{N}(u+z)- \mathrm{d}\widetilde{\bm{N}}(u+z) \big|/ \mathrm{d} u \preceq  2 v_1 \left( \bm{\Omega}^{n} \bm{J}+ \sum_{i=1}^{n} \bm{\Omega}^{i-1} \bm{\eta} \bm{J}\right).
\end{equation*}
We have thus completed the induction for $ \mathbb{E}\big| \mathrm{d} \bm{N}(u+z)- \mathrm{d} \widetilde{\bm{N}}(u+z) \big|/ \mathrm{d} u$.
To summarize, for any $n$ and $u > (n-1)b$, it holds that
\begin{equation*}
\mathbb{E}\big|\mathrm{d} \bm{N}(u+z)- \mathrm{d}\widetilde{\bm{N}}(u+z) \big|/ \mathrm{d} u \preceq  2 v_1 \left( \bm{\Omega}^{n} \bm{J}+  \sum_{i=1}^{n} \bm{\Omega}^{i-1} \bm{\eta}_{j,\cdot} \bm{J}\right).
\end{equation*}
Equivalently, we can say that for any $u \in \mathbb{R}^+$, it holds that
\begin{equation*}
\mathbb{E}\big|\mathrm{d} \bm{N}(u+z)- \mathrm{d}\widetilde{\bm{N}}(u+z) \big|/ \mathrm{d} u \preceq  2 v_1 \left(  \bm{\Omega}^{\lfloor u/b +1 \rfloor} \bm{J}+  \sum_{i=1}^{\lfloor u/b +1 \rfloor} \bm{\Omega}^{i-1}\bm{\eta} \bm{J}\right),
\end{equation*}
as required. \QEDB

\subsection{Proof of Lemma~\ref{lmm::bound_crosscovariance}}\label{sec::proof_crosscovariance}
As shown in the proof of Lemma~\ref{lmm::bound_expectation}, we know that the limiting processes $\bm{N}$ and $\widetilde{\bm{N}}$ exist and satisfy \eqref{eqn::N_infinity} and \eqref{eqn::Ntilde_infinity}, respectively.


We first bound the cross-term  $ \mathbb{E}\big| \mathrm{d} \bm{N}(t') \mathrm{d} \bm{N}^{\T}(u+z)- \mathrm{d} \widetilde{\bm{N}}(t') \mathrm{d} {\bm{N}}^{\T}(u+z)\big|/(\mathrm{d} t' \mathrm{d}u)$.
To begin, we show that there exists an upper bound on $ \mathbb{E}\big| \mathrm{d} \bm{N}(t') \mathrm{d} \bm{N}^{\T}(u+z)- \mathrm{d} \widetilde{\bm{N}}(t') \mathrm{d} {\bm{N}}^{\T}(u+z)\big|/(\mathrm{d} t' \mathrm{d}u)$ for any $u \in \mathbb{R}$ and $t'\geq u+z$.
The triangle inequality yields that
\begin{equation}\label{eqn::cross_term_zero}
\begin{aligned}
& \mathbb{E}\big| \mathrm{d} \bm{N}(t') \mathrm{d} \bm{N}^{\T}(u+z)- \mathrm{d} \widetilde{\bm{N}}(t') \mathrm{d} {\bm{N}}^{\T}(u+z)\big|/(\mathrm{d} t' \mathrm{d}u)\\
\preceq & \mathbb{E}\big| \mathrm{d} \bm{N}(t') \mathrm{d} \bm{N}^{\T}(u+z)\big|/(\mathrm{d} t' \mathrm{d}u) + \mathbb{E} \big| \mathrm{d} \widetilde{\bm{N}}(t') \mathrm{d} {\bm{N}}^{\T}(u+z)\big|/(\mathrm{d} t' \mathrm{d}u)\\
= & \bm{V}(t'-u-z) + \bm{\Lambda} \bm{\Lambda}^{\T} + \mathbb{E} \big| \mathrm{d} \widetilde{\bm{N}}(t') \mathrm{d} {\bm{N}}^{\T}(u+z)\big|/(\mathrm{d} t' \mathrm{d}u),
\end{aligned}
\end{equation}
where the equality follows from the definition of $\bm{V}$ and $\bm{\Lambda}$.

To find an upper bound for $\mathbb{E} \big| \mathrm{d} \widetilde{\bm{N}}(t') \mathrm{d} {\bm{N}}^{\T}(u+z)\big|/(\mathrm{d} t' \mathrm{d}u)$, recall the construction of $\{\bm{N}^{(i)}\}_{i=1}^{\infty}$: For $i=1$, we have
\begin{equation}
\frac{\mathbb{E} \big| \mathrm{d} \widetilde{{N}}^{(1)}_j(t') \mathrm{d} {{N}}^{(1)}_k(u+z)\big|}{\mathrm{d} t' \mathrm{d}u} =  \frac{\mathbb{E} \big|  \widetilde{{N}}^{(0)}_j\big([0,\phi(\mu_j)] \times \mathrm{d} t'\big) {{N}}^{(0)}_k \big([0,\phi(\mu_k)] \times \mathrm{d} (u+z)  \big) \big|}{\mathrm{d} t' \mathrm{d}u}=\phi_j(\mu_j)\phi_k(\mu_k),
\end{equation}
where the first equality follows by definition and the second equality follows from the fact that $\widetilde{{N}}^{(0)}_j$ and ${{N}}^{(0)}_k$ are independent when $t'\geq u + z > z$.

Define $\bm{A}^{(1)}=\big(A^{(1)}_{j,k}\big)$ where  $A^{(1)}_{j,k}=\phi_j(\mu_j)\phi_k(\mu_k)$.
Suppose that  for $m=i$
\begin{equation}\label{eqn::crossterm}
{\mathbb{E} \big| \mathrm{d} \widetilde{\bm{N}}^{(i)}(t') \mathrm{d} {\bm{N}}^{(i)}(u+z)\big|}/{\mathrm{d} t' \mathrm{d}u} \preceq \bm{A}^{(i)}.
\end{equation}
We can see that \eqref{eqn::crossterm} holds for $i=1$.

Then for $m=i+1$, it follows that
\begin{equation*}
\begin{aligned}
& {\mathbb{E} \big| \mathrm{d} \widetilde{{N}}_j^{(i+1)}(t') \mathrm{d} {{N}}_k^{(i+1)}(u+z)\big|}/{\mathrm{d} t' \mathrm{d}u}\\
= & \mathbb{E} \big|  \widetilde{{N}}^{(0)}_j\big(\big[0,\lambda_j^{(i+1)}(t') \big] \times \mathrm{d} t'\big) {{N}}^{(0)}_k\big(\big[0,\lambda_{k}^{(i+1)}(u+z)\big] \times \mathrm{d} (u+z)\big)\big|/{\mathrm{d} t' \mathrm{d}u} \\
= & \mathbb{E} \Big|  \phi_j\Big\{  \mu_j + \sum_{l=1}^p \int_0^{\infty} \omega_{l,j}(\Delta) \mathrm{d} \widetilde{N}_l^{(i)}(t'- \Delta)  \Big\} \phi_k\Big\{  \mu_j + \sum_{l=1}^p \int_0^{\infty} \omega_{l,j}(\Delta) \mathrm{d} {N}_l^{(i)}(u+z- \Delta)  \Big\} \Big|,
\end{aligned}
\end{equation*}
where the equalities follow by definition.

Using the Lipschitz condition for the link function yields
\begin{equation}\label{eqn::lip_cond}
\begin{aligned}
 \phi_j\left\{ \mu_j + \sum_{l=1}^p \int_0^{\infty} \omega_{l,j}(\Delta) \mathrm{d} \widetilde{N}_l^{(i)}(t'- \Delta)  \right\} \leq & \phi_j(\mu_j)+ \left| \sum_{l=1}^p \int_0^{\infty} \omega_{l,j}(\Delta) \mathrm{d} \widetilde{N}_l^{(i)}(t'- \Delta)\right|
\end{aligned}
\end{equation}
With this, we see that
\begin{equation}\label{eqn::prod_links}
\begin{aligned}
& \mathbb{E} \Big|  \phi_j\Big\{ \mu_j + \sum_{l=1}^p \int_0^{\infty} \omega_{l,j}(\Delta) \mathrm{d} \widetilde{N}_l^{(i)}(t'- \Delta)  \Big\}  \phi_k\Big\{ \mu_k + \sum_{l=1}^p \int_0^{\infty} \omega_{l,k}(\Delta) \mathrm{d} {N}_l^{(i)}(u+z- \Delta)  \Big\} \Big|\\
\leq & \mathbb{E} \Big|\Big[ \phi_j(\mu_j)+ \sum_{l=1}^p \int_0^{\infty} \big|\omega_{l,j}(\Delta)\big| \mathrm{d} \widetilde{N}_l^{(i)}(t'- \Delta)  \Big] \Big[ \phi_k(\mu_k) + \sum_{l=1}^p \int_0^{\infty} \big|\omega_{l,k}(\Delta)\big| \mathrm{d} {N}_l^{(i)}(u+z- \Delta) \Big] \Big|\\
= & \mathbb{E} \Big| \phi_j(\mu_j)\phi_k(\mu_k) +  \phi_k(\mu_k)  \sum_{l=1}^p \int_0^{\infty} \big|\omega_{l,j}(\Delta)\big| \mathrm{d} \widetilde{N}_l^{(i)}(t'- \Delta) + \\
& \phi_j(\mu_j)\sum_{l=1}^p \int_0^{\infty} \big|\omega_{l,k}(\Delta)\big| \mathrm{d} {N}_l^{(i)}(u+z- \Delta) + \\
&  \sum_{l=1}^p \int_0^{\infty} \big|\omega_{l,j}(\Delta)\big| \mathrm{d} \widetilde{N}_l^{(i)}(t'- \Delta) \sum_{l'=1}^p \int_0^{\infty} \big|\omega_{l',j}(\Delta)\big| \mathrm{d} {N}_{l'}^{(i)}(u+z- \Delta)\Big|\\
=  & C^{(i)}_{j,k} +\mathbb{E} \Big|  \sum_{l=1}^p \int_0^{\infty} \big|\omega_{l,j}(\Delta)\big| \mathrm{d} \widetilde{N}_l^{(i)}(t'- \Delta) \sum_{l'=1}^p \int_0^{\infty} \big|\omega_{l',j}(\Delta)\big| \mathrm{d} {N}_{l'}^{(i)}(u+z- \Delta)\Big|
\end{aligned}
\end{equation}
where
\begin{equation*}
\begin{aligned}
C^{(i)}_{j,k} \equiv & \mathbb{E} \Big| \phi_j(\mu_j)\phi_k(\mu_k) +  \phi_k(\mu_k)  \sum_{l=1}^p \int_0^{\infty} \big|\omega_{l,j}(\Delta)\big| \mathrm{d} \widetilde{N}_l^{(i)}(t'- \Delta)\\
& +\phi_j(\mu_j)\sum_{l'=1}^p \int_0^{\infty} \big|\omega_{l',j}(\Delta)\big| \mathrm{d} {N}_{l'}^{(i)}(u+z- \Delta)\Big|\\
\leq & \phi_j(\mu_j)\phi_k(\mu_k)+ \phi_k(\mu_k)\bm{\Omega}_{j,\cdot} \cdot \bm{\Lambda}+\phi_j(\mu_j)\bm{\Omega}_{k,\cdot} \cdot \bm{\Lambda}
\end{aligned}
\end{equation*}
is bounded by construction.
Rewriting the right-hand side of \eqref{eqn::prod_links} yields
\begin{equation*}
\begin{aligned}
& \mathbb{E} \Big|  \phi_j\Big\{ \mu_j + \sum_{l=1}^p \int_0^{\infty} \omega_{l,j}(\Delta) \mathrm{d} \widetilde{N}_l^{(i)}(t'- \Delta)  \Big\}  \phi_k\Big\{ \mu_k + \sum_{l=1}^p \int_0^{\infty} \omega_{l,k}(\Delta) \mathrm{d} {N}_l^{(i)}(u+z- \Delta)  \Big\} \Big|\\
\leq & C^{(i)}_{j,k} + \mathbb{E} \Big| \sum_{ 1\leq l,l' \leq p} \int \int \big|\omega_{l,j}(\Delta) \omega_{l',k}(\Delta')\big| \mathrm{d}\widetilde{N}_l^{(i)}(t'-\Delta) \mathrm{d}N_{l'}^{(i)}(u+z-\Delta)\Big| \\
= & C^{(i)}_{j,k} +  \sum_{ 1\leq l,l' \leq p} \int \int \big|\omega_{l,j}(\Delta) \omega_{l',k}(\Delta')\big| \mathbb{E} \Big| \mathrm{d}\widetilde{N}_l^{(i)}(t'-\Delta) \mathrm{d}N_{l'}^{(i)}(u+z-\Delta)\Big|\\
\leq & C^{(i)}_{j,k} +  \sum_{ 1\leq l,l' \leq p} \int \int \big|\omega_{l,j}(\Delta) \omega_{l',k}(\Delta')\big| A^{(i)}_{l,l'} \mathrm{d} \Delta \mathrm{d} \Delta' \\
\leq & C^{(i)}_{j,k} + \sum_{l,l'} \Omega_{j,l} \Omega_{k,l'} A^{(i)}_{l,l'}\\
= & C^{(i)}_{j,k} + \bm{\Omega}_{j,\cdot} \bm{A}^{(i)} \bm{\Omega}_{k, \cdot}^{\T},
\end{aligned}
\end{equation*}
where the second equality holds since all terms in the integral are non-negative, and the second-to-last inequality holds due to the induction condition \eqref{eqn::crossterm}.

Therefore, we can see that \eqref{eqn::crossterm} holds for $i+1$ with $\bm{A}^{(i+1)}\equiv  \bm{\Omega}^{\T} \bm{A}^{(i)} \bm{\Omega} + \bm{C}^{(i)}$.
By induction,
$$\bm{A}^{(i+1)} = \left(\bm{\Omega}^{\T} \right)^{i} \bm{A}^{(1)} \bm{\Omega}^{i} +\sum_{m=1}^i \left(\bm{\Omega}^{\T} \right)^{i-m} \bm{C}^{(m)} \bm{\Omega}^{i-m}.$$
Given that $\Gamma_{\max}(\bm{\Omega}) <1$ from Assumption~\ref{asmp::spectralradius}, we know that $\bm{A}^{(\infty)}\equiv \lim_{i\rightarrow \infty} A^{(i)}$ exists and thus $\mathbb{E} \big| \mathrm{d} \widetilde{\bm{N}}(t') \mathrm{d} {\bm{N}}^{\T}(u+z)\big|/(\mathrm{d} t' \mathrm{d}u)$ has a well-defined element-wise bound, which we denote by $\bm{A}$.

We now prove the inequality in \eqref{eqn::cross_term_zero}. First, for any $u$,
\begin{equation}\label{eqn::cross_term_zero_cmp}
\begin{aligned}
& \mathbb{E}\big| \mathrm{d} \bm{N}(t') \mathrm{d} \bm{N}^{\T}(u+z)- \mathrm{d} \widetilde{\bm{N}}(t') \mathrm{d} {\bm{N}}^{\T}(u+z)\big|/(\mathrm{d} t' \mathrm{d}u)\\
\preceq &  \bm{V}(t'-u-z) + \bm{\Lambda} \bm{\Lambda}^{\T} + \bm{A} \preceq v_2 \bm{J} \bm{J}^{\T},
\end{aligned}
\end{equation}
where $v_2=\max_{j,k,\Delta} \big\{|V_{j,k}(\Delta)| +\Lambda_j \Lambda_k + A_{j,k}\big\}$.
Suppose, for $m=n-1$, that, for $u \geq (n-2)b$,
$$ \mathbb{E}\big| \mathrm{d} \bm{N}(t') \mathrm{d} \bm{N}^{\T}(u+z)- \mathrm{d} \widetilde{\bm{N}}(t') \mathrm{d} {\bm{N}}^{\T}(u+z)\big|/(\mathrm{d} t' \mathrm{d}u) \preceq v_2\left( \bm{\Omega}^{n-1} \bm{J}\bm{J}^{\T} +  \sum_{i=1}^{n-1} \bm{\Omega}^{i-1} \bm{\eta} \bm{J}\bm{J}^{\T}\right),$$
where $\bm{\Omega}^{0}$ is the identity matrix.
Then, for $m=n$ (i.e., $u \geq (n-1)b$),
\begin{equation*}
\begin{aligned}
&  \mathbb{E}\big| \mathrm{d} {N}_j(t') \mathrm{d}{N}_{k}(u+z)- \mathrm{d} \widetilde{{N}}_j(t') \mathrm{d} {{N}}_{k}(u+z)\big|/(\mathrm{d} t' \mathrm{d}u)\\
= & \mathbb{E}\big|
\big[\tilde{\lambda}_j(t')-\lambda_k(u+z) \big] \mathrm{d} {{N}}_{k}(u+z)\big|/\mathrm{d}u\\
= & \mathbb{E}\left| \left[
\phi\Big( \mu_j + \bm{\omega}_{\cdot,j}* \mathrm{d} \widetilde{\bm{N}}(t')  \Big) -  \phi\Big( \mu_j + \bm{\omega}_{\cdot,j} * \mathrm{d} {\bm{N}}(u+z)  \Big)
\right] \mathrm{d} {{N}}_{k}(u+z)\right|/\mathrm{d}u\\
\leq & \mathbb{E}\left| \Big|
\phi\Big( \mu_j + \bm{\omega}_{\cdot,j}* \mathrm{d} \widetilde{\bm{N}}(t')  \Big) -  \phi\Big( \mu_j + \bm{\omega}_{\cdot,j} * \mathrm{d} {\bm{N}}(u+z)  \Big)
\Big| \mathrm{d} {{N}}_{k}(u+z)\right|/\mathrm{d}u\\
\leq & \mathbb{E}\left| \Big|\sum_{l=1}^p \int_0^{\infty} \omega_{l,j}(\Delta) \big[ \mathrm{d}\widetilde{N}_l(t'-\Delta) - \mathrm{d}{N}_l(u+z-\Delta) \big]  \Big| \mathrm{d} {{N}}_{k}(u+z)\right|/\mathrm{d}u\\
\leq & \sum_{l=1}^p \int_0^{\infty} \mathbb{E}\big| \omega_{l,j}(\Delta) \big[ \mathrm{d}\widetilde{N}_l(t'-\Delta) - \mathrm{d}{N}_l(u+z-\Delta) \big]  \big| \mathrm{d} {{N}}_{k}(u+z)/\mathrm{d}u,
\end{aligned}
\end{equation*}
where the second inequality follows from the Lipschitz condition of $\phi_j$ \eqref{eqn::lip_cond}.

Then, for each $l$, we can see that
\begin{equation*}
\begin{aligned}
& \int_0^{\infty} \mathbb{E}\big| \omega_{l,j}(\Delta) \big[ \mathrm{d}\widetilde{N}_l(t'-\Delta) - \mathrm{d}{N}_l(u+z-\Delta) \big]  \big| \mathrm{d} {{N}}_{k}(u+z)/\mathrm{d}u\\
\leq &  \int_0^{\infty} \big|\omega_{l,j}(\Delta) \big|\mathbb{E} \big|  \mathrm{d}\widetilde{N}_l(t'-\Delta)\mathrm{d} {{N}}_{k}(u+z) - \mathrm{d}{N}_l(u+z-\Delta)\mathrm{d} {{N}}_{k}(u+z)  \big|/\mathrm{d}u\\
= &  \left\{ \int_0^{b} + \int_{b}^{\infty}\right\}\big|\omega_{l,j}(\Delta)  \big| \,  \mathbb{E} \big|  \mathrm{d}\widetilde{N}_l(t'-\Delta)\mathrm{d} {{N}}_{k}(u+z) - \mathrm{d}{N}_l(u+z-\Delta)\mathrm{d} {{N}}_{k}(u+z)  \big|/\mathrm{d}u\\
\leq & v_2 \Omega_{j,l} \Big[ \bm{\Omega}^{n-1} \bm{J}\bm{J}^{\T} +  \sum_{i=1}^{n-1} \bm{\Omega}^{i-1} \bm{\eta} \bm{J}\bm{J}^{\T} \Big]_{l,k} +\eta_{j,l} v_2.
\end{aligned}
\end{equation*}
Thus,
\begin{equation*}
\begin{aligned}
&  \mathbb{E}\big| \mathrm{d} {N}_j(t') \mathrm{d}{N}_{k}(u+z)- \mathrm{d} \widetilde{{N}}_j(t') \mathrm{d} {{N}}_{k}(u+z)\big|/(\mathrm{d} t' \mathrm{d}u) \\
\leq & v_2\bm{\Omega}_{j,\cdot}^{\T}\Big[  \bm{\Omega}^{n-1} \bm{J}\bm{J}^{\T} +  \sum_{i=1}^{n-1} \bm{\Omega}^{i-1} \bm{\eta} \bm{J}\bm{J}^{\T} \Big]_{\cdot,k}+   v_2 \eta_{j,l}.
\end{aligned}
\end{equation*}
Therefore, we have shown that, for $u \geq (n-1)b$
\begin{equation}\label{eqn::cross_term}
\mathbb{E}\big| \mathrm{d} \bm{N}(t') \mathrm{d} \bm{N}^{\T}(u+z)- \mathrm{d} \widetilde{\bm{N}}(t') \mathrm{d} {\bm{N}}^{\T}(u+z)\big|/(\mathrm{d} t' \mathrm{d}u) \preceq  v_2\left( \bm{\Omega}^{n} \bm{J}\bm{J}^{\T} +   \sum_{i=1}^{n} \bm{\Omega}^{i-1} \bm{\eta} \bm{J}\bm{J}^{\T}\right).
\end{equation}

Finally, we return to bounding $\mathbb{E}\big| \mathrm{d} \bm{N}(t') \mathrm{d} \bm{N}^{\T}(u+z)- \mathrm{d} \widetilde{\bm{N}}(t') \mathrm{d} \widetilde{\bm{N}}^{\T}(u+z)\big|/(\mathrm{d} t' \mathrm{d}u)$.
Note that when $t'=u+z$,
$$  \mathbb{E}\big| \mathrm{d} {N}_k(u+z) \mathrm{d} {N}_k(u+z)- \mathrm{d}  \widetilde{{N}}_k(u+z) \mathrm{d} \widetilde{{N}}_k(u+z)\big| = \mathbb{E}\big| \mathrm{d} {N}_k(u+z)- \mathrm{d} \widetilde{{N}}_k(u+z)\big|. $$
This quantity was bounded in Lemma~\ref{lmm::bound_expectation}.
Moreover,
\begin{equation*}
\small
\mathbb{E}\big| \mathrm{d} \bm{N}(t') \mathrm{d}\bm{N}^{\T}(u+z)-d  \widetilde{\bm{N}}(t')\mathrm{d} \widetilde{\bm{N}}^{\T}(u+z)\big| = \left[ \mathbb{E}\big| \mathrm{d}\bm{N}(u+z) \mathrm{d} \bm{N}^{\T}(t')-\mathrm{d}  \widetilde{\bm{N}}(u+z) \mathrm{d} \widetilde{\bm{N}}^{\T}(t')\big| \right]^{\T}.
\end{equation*}
Hence, it suffices to consider the case when $t' > u+z$.

Now, for any $u \geq (n-1)b$ and $t' > u+z$, we have
\begin{equation*}
\begin{aligned}
& \mathbb{E}\big| \mathrm{d} {N}_j(t') \mathrm{d} {N}_k^{\T}(u+z)- \mathrm{d} \widetilde{{N}}_j(t') \mathrm{d} \widetilde{{N}}_k^{\T}(u+z)\big|/(\mathrm{d} t' \mathrm{d}u) \\
= & \mathbb{E}\big| \mathrm{d} {N}_j(t') \mathrm{d} {N}_k^{\T}(u+z)-  \mathrm{d} \widetilde{{N}}_j(t') \mathrm{d} {{N}}_k^{\T}(u+z)+  \mathrm{d} \widetilde{{N}}_j(t') \mathrm{d} {{N}}_k^{\T}(u+z)- \mathrm{d} \widetilde{{N}}_j(t') \mathrm{d} \widetilde{{N}}_k^{\T}(u+z)\big|/(\mathrm{d} t' \mathrm{d}u)\\
\leq & \mathbb{E}\big| \big[\mathrm{d} {N}_j(t') -  \mathrm{d} \widetilde{{N}}_j(t')\big] \mathrm{d} {{N}}_k^{\T}(u+z) \big| /(\mathrm{d} t' \mathrm{d}u) +  \mathbb{E}\big| \mathrm{d} \widetilde{{N}}_j(t') \big[\mathrm{d} {{N}}_k^{\T}(u+z)-  \mathrm{d} \widetilde{{N}}_k^{\T}(u+z)\big]\big|/(\mathrm{d} t' \mathrm{d}u)\\
= &  \mathbb{E}\big| \mathbb{E}\big\{ \big[\mathrm{d} {N}_j(t') -  \mathrm{d} \widetilde{{N}}_j(t')\big]/ \mathrm{d} t'  \mid \mathcal{H}_{t'}, \widetilde{\mathcal{H}}_{t'}\big\}\mathrm{d} {{N}}_k(u+z) \big| /(\mathrm{d}u) + \\
 & \mathbb{E}\big| \mathbb{E}\big\{ \big[ \mathrm{d} \widetilde{{N}}_j(t')\big]/ \mathrm{d} t'  \mid   \widetilde{\mathcal{H}}_{t'}\big\}\big[\mathrm{d} {{N}}_k(u+z)-  \mathrm{d} \widetilde{{N}}_k(u+z)\big] \big|/(\mathrm{d}u)\\
= &  \mathbb{E}\big|\big[ \lambda_j(t') -  \tilde{\lambda}_j(t')\big] \mathrm{d} {{N}}_k(u+z) \big| /(\mathrm{d}u) +  \mathbb{E}\big| \tilde{\lambda}_j(t') \big[\mathrm{d} {{N}}_k(u+z)-  \mathrm{d} \widetilde{{N}}_k(u+z)\big] \big|/(\mathrm{d}u).
\end{aligned}
\end{equation*}

Next, we use the Lipschitz condition of the link function $\phi_j$.
Recall that
$$ \lambda_j(t) = \phi_j\Big\{ \mu_j + \big(\bm{\omega}_{\cdot,j}* \mathrm{d} {\bm{N}}\big)(t)  \Big\} \quad  {\rm and} \quad   \tilde{\lambda}_j(t) = \phi_j\Big\{ \mu_j + \big(\bm{\omega}_{\cdot,j}* \mathrm{d} \widetilde{\bm{N}}\big)(t)  \Big\}.$$
Then,
\begin{equation*}
\begin{aligned}
& \mathbb{E}\big|\big[ \lambda_j(t') -  \tilde{\lambda}_j(t')\big] \mathrm{d} {{N}}_k(u+z) \big| /(\mathrm{d}u) \\
\leq & \mathbb{E}\Big|\Big\{\sum_{l=1}^p \int_0^{\infty} \omega_{l,j}(\Delta) \big[\mathrm{d} N_l(t'-\Delta) -\mathrm{d} \widetilde{N}_l(t'-\Delta)\big] \Big\} \mathrm{d} {{N}}_k(u+z) \Big| /(\mathrm{d}u)\\
\leq & \sum_{l=1}^p \int_0^{\infty} \big|\omega_{l,j}(\Delta)\big| \mathrm{d} \Delta \max_{u \geq (n-1)b}\Big\{ \frac{\mathrm{d} N_l(t'-\Delta) \mathrm{d} {{N}}_k(u+z) -\mathrm{d} \widetilde{N}_l(t'-\Delta) \mathrm{d} {{N}}_k(u+z)}{\mathrm{d}\Delta  \mathrm{d}u }  \Big\}\\
\leq & v_2\sum_{l=1}^p \Omega_{j,l} \Big[ \bm{\Omega}^{n} \bm{J}\bm{J}^{\T} +   \sum_{i=1}^{n} \bm{\Omega}^{i-1}\bm{\eta}\bm{J}\bm{J}^{\T}\Big]_{l,k}.
\end{aligned}
\end{equation*}
And, similarly,
\begin{equation*}
\begin{aligned}
&\mathbb{E}\big| \tilde{\lambda}_j(t') \big[\mathrm{d} {{N}}_k(u+z)-  \mathrm{d} \widetilde{{N}}_k(u+z)\big] \big|/(\mathrm{d}u) \\
\leq &  \mathbb{E}\Big| \Big[\phi_j(\mu_j) + \sum_{l=1}^p \int_0^{\infty} \omega_{l,j}(\Delta)\mathrm{d} \widetilde{N}_l (t'-\Delta) \Big]  \big[\mathrm{d} {{N}}_k(u+z)-  \mathrm{d} \widetilde{{N}}_k(u+z)\big] \Big|/(\mathrm{d}u)\\
\leq & \phi_j(\mu_j)  \max_{u \geq (n-1)b} \mathbb{E}\big|\mathrm{d} {{N}}_k(u+z)-  \mathrm{d} \widetilde{{N}}_k(u+z)\big| /(\mathrm{d}u)+\\
&  \sum_{l=1}^p \int_0^{\infty} |\omega_{l,j}(\Delta)| \mathrm{d} \Delta \max_{u \geq (n-1)b} \left\{\frac{\mathbb{E}\big|\mathrm{d} \widetilde{N}_l (t'-\Delta)\mathrm{d} {{N}}_k(u+z)-  \mathrm{d} \widetilde{N}_l (t'-\Delta) \mathrm{d} \widetilde{{N}}_k(u+z) \big|}{\mathrm{d} \Delta \mathrm{d}u}  \right\}\\
\leq & 2v_1^2 \left( \bm{\Omega}^{n} \bm{J}\right)_k + 2v_1^2   \sum_{i=1}^{n} \left(\bm{\Omega}^{i-1} \bm{\eta}\bm{J}\right)_k+\sum_{l=1}^p \Omega_{j,l} v_2 \Big[\bm{\Omega}^{n} \bm{J}\bm{J}^{\T} + \sum_{i=1}^{n} \bm{\Omega}^{i-1} \bm{\eta} \bm{J}\bm{J}^{\T}\Big]_{l,k},
\end{aligned}
\end{equation*}
where the second inequality follows from the Lipschitz condition of $\phi_j$ \eqref{eqn::lip_cond}.

To summarize, we have shown that
\begin{equation*}
\begin{aligned}
& \mathbb{E}\big| \mathrm{d} {N}_j(t') \mathrm{d} {N}_k^{\T}(u+z)- \mathrm{d} \widetilde{{N}}_j(t') \mathrm{d} \widetilde{{N}}_k^{\T}(u+z)\big|/(\mathrm{d} t' \mathrm{d}u) \\
\leq & 2v_2 \bm{\Omega}_{j,\cdot} \Big[\bm{\Omega}^{n} \bm{J}\bm{J}^{\T} +  \sum_{i=1}^{n} \bm{\Omega}^{i-1} \bm{\eta}\bm{J}\bm{J}^{\T}]_{\cdot,k}+2v_1^2 \left\{  \left( \bm{\Omega}^{n} \bm{J}\right)_k +\sum_{i=1}^{n} \left( \bm{\Omega}^{i-1} \bm{\eta} \bm{J}\right)_k\right\}.
\end{aligned}
\end{equation*}
In the matrix form, we have, for $u > (n-1)b$
\begin{equation*}
\begin{aligned}
& \mathbb{E}\big| \mathrm{d} \bm{N}(t') \mathrm{d} \bm{N}^{\T}(u+z)- \mathrm{d} \widetilde{\bm{N}}(t') \mathrm{d} \widetilde{\bm{N}}^{\T}(u+z)\big|/(\mathrm{d} t' \mathrm{d}u) \\
\preceq &
2 v_2 \bm{\Omega}^{n+1} \bm{J}\bm{J}^{\T} +2 v_2^2  \bm{J}\bm{J}^{\T} \big(\bm{\Omega}^{n}\big)^{\T}  +2 v_2 \sum_{i=1}^{n} \bm{\Omega}^i \bm{\eta}\bm{J}\bm{J}^{\T} + 2 v_2^2 \sum_{i=1}^{n} \bm{\eta}\bm{J}\bm{J}^{\T} \big(\bm{\Omega}^{i}\big)^{\T},
\end{aligned}
\end{equation*}
or, alternatively,
\begin{equation}
\begin{aligned}
& {\mathbb{E}\big| \mathrm{d} \widetilde{\bm{N}}(t')   \mathrm{d} \widetilde{\bm{N}}(z+u)  -  \mathrm{d} {\bm{N}}(t') \mathrm{d} {\bm{N}}(z+u) \big|}/\big({\mathrm{d} u \mathrm{d} t'} \big) \\
\preceq  & 2 v_2 \left\{ \bm{\Omega}^{\lfloor u/b +2 \rfloor} \bm{J}\bm{J}^{\T} + \sum_{i=1}^{\lfloor u/b +1 \rfloor} \bm{\Omega}^i \bm{\eta} \bm{J}\bm{J}^{\T}\right\} + 2 v_1^2 \left\{  \bm{J}\bm{J}^{\T} \big(\bm{\Omega}^{\lfloor u/b +1 \rfloor}\big)^{\T}  + \sum_{i=1}^{\lfloor u/b +1 \rfloor} \bm{\eta} \bm{J}\bm{J}^{\T} \big(\bm{\Omega}^{i}\big)^{\T}\right\},
\end{aligned}
\end{equation}
as required. \QEDB

\subsection{Proof of Lemma~\ref{lmm::product}}\label{sec::proof_product}

	Let $M= (nK_2 /K_1)^{1/2} $.
	We have that
	\begin{align*}
	P(|Z_1 Z_2|> n ) & = P( \{ |Z_2| \geq M, |Z_1 Z_2|> n  \} \cup  \{ |Z_2| < M, |Z_1 Z_2|> n  \}  ) \\
	& = P( \{ |Z_2| \geq M, |Z_1 Z_2|> n  \} ) + P( \{ |Z_2| < M, |Z_1 Z_2|> n  \}  ) \\
	& \leq P( |Z_2| \geq M ) + P( \{ |Z_2| < M, M |Z_1|> n  \}  ) \\
	& \leq P( |Z_2| \geq M ) + P(M |Z_1|> n ) \\
	&  \leq 2\exp\big( 1- (n/K_1 K_2)^{1/2} \big)\\
	& \leq \exp\big( 1- (n/K^*)^{1/2}\big).
	\end{align*}
	where the last inequality follows from the proof of Lemma A.2 in \cite{fan2014}. \QEDB

\subsection{Proof of Lemma~\ref{lmm::elementwise_expectation}}\label{sec::proof_elementwise_expectation}
From Assumption~\ref{asmp::uniform}, we have that  $(\bm{\Omega}\bm{J})_{j} \leq \rho_{\Omega}$, which is equivalent to
$ \bm{\Omega}\bm{J}\preceq \rho_{\Omega}\bm{J}$.
Then, by induction, $\bm{\Omega}^n\bm{J}\preceq \rho_{\Omega}^n\bm{J}$.
Hence, from Theorem~\ref{thm::coupling}, we know that,  for each $j$,
\begin{equation*}
\begin{aligned}
\mathbb{E}\big|\mathrm{d} \widetilde{{N}}_j(z+u) - \mathrm{d}{{N}}_j(z+u) \big|/\mathrm{d}u  \leq  & \big( 2 v \bm{\Omega}^{\lfloor u/b +1 \rfloor} \bm{J}+ 2v  \sum_{i=1}^{\lfloor u/b +1 \rfloor} \bm{\Omega}^{i-1} \bm{\eta}\bm{J}\big)_j\\
= & \big( 2 v \bm{\Omega}^{\lfloor u/b +1 \rfloor} \bm{J}\big)_j+ 2v \sum_{i=1}^{\lfloor u/b +1 \rfloor} \big(\bm{\Omega}^{i-1} \bm{\eta}\bm{J}\big)_j\\
\leq & \big( 2 v \bm{\Omega}^{\lfloor u/b +1 \rfloor} \bm{J}\big)_j+ 2v \max_{j} [\bm{\eta}_{j,\cdot} \bm{J}] \sum_{i=1}^{\lfloor u/b +1 \rfloor} \big(\bm{\Omega}^{i-1} \bm{J}\big)_j\\
\leq & 2v \rho_{\Omega}^{\lfloor u/b +1 \rfloor} + 2v \max_{j} [\bm{\eta}_{j,\cdot} \bm{J}]  \sum_{i=1}^{\lfloor u/b +1 \rfloor} \rho_{\Omega}^{i-1}\\
\leq & 2v \rho_{\Omega}^{\lfloor u/b +1 \rfloor} + 2v \max_{j} [\bm{\eta}_{j,\cdot} \bm{J}]  \frac{\rho_{\Omega}}{1-\rho_{\Omega}},
\end{aligned}
\end{equation*}
where we use that $v=\max(v_1,v_1^2,v_2)$.

Let $b=\max(b_0, \log^{1/(r+1)}(\rho_{\Omega}^{-1})(\constantTenP)^{-1/(r+1)} u^{1/(r+1)})$. Then by Assumption~\ref{asmp::tail_highd}
 $$\max_{j} [\bm{\eta}_{j,\cdot} \bm{J}]\leq  \constantElevenP \exp\{- \log^{r/(r+1)}(\rho_{\Omega}^{-1})(\constantTenP) ^{1/(r+1)}  u^{r/(r+1)}  \}.$$
Therefore,
\begin{equation*}
\begin{aligned}
& \mathbb{E}\big|\mathrm{d} \widetilde{{N}}_j(z+u) - \mathrm{d}{{N}}_j(z+u) \big|/\mathrm{d}u  \\
\leq & 2v \rho_{\Omega}^{\lfloor u/b +1 \rfloor} + 2v p \eta   \frac{\rho_{\Omega}}{1-\rho_{\Omega}}\\
\leq & [2 v  +2v \frac{\rho_{\Omega}}{1-\rho_{\Omega}}\constantElevenP ] \exp\{- \log^{r/(r+1)}(\rho_{\Omega}^{-1})(\constantTenP)^{1/(r+1)}  u^{r/(r+1)} \}\\
\equiv & \constantTwelveP v \exp\big(-\constantThirteenP u^{r/(r+1)}\big),
\end{aligned}
\end{equation*}
where we use the fact that $\lfloor u/b +1 \rfloor\geq u/b$.  \QEDB

\subsection{Proof of Lemma~\ref{lmm::elementwise_crosscovariance}}\label{sec::proof_elementwise_crosscovariance}


Similar to the proof of Lemma~\ref{lmm::elementwise_expectation}, we can rewrite the bound in Theorem~\ref{thm::coupling} as
\begin{equation}
\begin{aligned}
&  \mathbb{E}\big| \mathrm{d} \widetilde{{N}}_j(t')   \mathrm{d} \widetilde{{N}}_k(z+u)  -  \mathrm{d} {{N}}_j(t')\mathrm{d} {{N}}_k(z+u) \big|/(\mathrm{d} t' \mathrm{d} u) \\
\leq  & \left[ 2 v_2 \bm{\Omega}^{\lfloor u/b +2 \rfloor} \bm{J}\bm{J}^{\T} +2 v_1^2  \bm{J}\bm{J}^{\T} \big(\bm{\Omega}^{\lfloor u/b +1 \rfloor}\big)^{\T}  +2 v_2\sum_{i=1}^{\lfloor u/b +1 \rfloor} \bm{\Omega}^i \bm{\eta} \bm{J}\bm{J}^{\T} + 2 v^2_1\sum_{i=1}^{\lfloor u/b +1 \rfloor} \bm{\eta}\bm{J}\bm{J}^{\T} \big(\bm{\Omega}^{i}\big)^{\T}\right]_{j,k}\\
\leq &  2 v \left[ \bm{\Omega}^{\lfloor u/b +2 \rfloor}\bm{J}\right]_{j}  +2 v  \left[\bm{\Omega}^{\lfloor u/b +1 \rfloor} \bm{J}\right]_{k}  +2  v \sum_{i=1}^{\lfloor u/b +1 \rfloor} \left[ \bm{\Omega}^i \bm{\eta}\bm{J}\right]_{j}\\
\leq &  2 v \rho_{\Omega}^{\lfloor u/b +2 \rfloor}  +2 v \rho_{\Omega}^{\lfloor u/b +1 \rfloor}  +2  v \max_{j} [\bm{\eta}_{j,\cdot} \bm{J}] \sum_{i=1}^{\lfloor u/b +1 \rfloor} \rho_{\Omega}^i\\
\leq & 2 v \rho_{\Omega}^{\lfloor u/b +2 \rfloor}  +2 v  \rho_{\Omega}^{\lfloor u/b +1 \rfloor}  +2  v\max_{j} [\bm{\eta}_{j,\cdot} \bm{J}] \frac{\rho_{\Omega}}{1-\rho_{\Omega}},
\end{aligned}
\end{equation}
where we use  $v=\max(v_1,v_1^2,v_2)$ in the second inequality, and Assumption~\ref{asmp::tail_highd} in the second-to-last inequality.

Recall that we set $b=\max(b_0, \log^{1/(r+1)}(\rho_{\Omega}^{-1})(\constantTenP)^{-1/(r+1)} u^{1/(r+1)})$. And thus Assumption~\ref{asmp::tail_highd} yields that
$$\max_{j} [\bm{\eta}_{j,\cdot} \bm{J}] \leq \constantElevenP \exp\{- \log^{r/(r+1)}(\rho_{\Omega}^{-1})(\constantTenP)^{1/(r+1)}  u^{r/(r+1)}  \}.$$
Thus, using the fact that $\rho_{\Omega}<1$,
\begin{equation}
\begin{aligned}
&  \mathbb{E}\big| \mathrm{d} \widetilde{{N}}_j(t')   \mathrm{d} \widetilde{{N}}_k(z+u)  -  \mathrm{d} {{N}}_j(t')\mathrm{d} {{N}}_k(z+u) \big|/(\mathrm{d} t' \mathrm{d} u) \\
\leq & \left(2v+2v+ 2v \frac{\rho_{\Omega}}{1-\rho_{\Omega}}+2p^{3/2}v \frac{\rho_{\Omega}}{1-\rho_{\Omega}} \right)\exp\left\{- \log^{r/(r+1)}(\rho_{\Omega}^{-1})(\constantTenP)^{1/(r+1)}  u^{r/(r+1)} \right\}\\
\equiv & \constantFourteenP v\exp\big(-\constantFifteenP u^{r/(r+1)}\big).
\end{aligned}
\end{equation}
\QEDB

\subsection{Proof of Lemma~\ref{lmm::exponential_tail}}\label{sec::proof_exp}

We discuss two scenarios depending on which condition holds in Assumption~\ref{asmp::bounded}.

(i) Suppose that Assumption~\ref{asmp::bounded}a holds, i.e., $\lambda_j(t) \leq \phi_{\max}$. We construct a dominating process $\widehat{\bm{N}}$ as
\begin{equation}\label{eqn::dominating_process}
\mathrm{d} \widehat{N}_j(t)  = {N}^{(0)}_j\big( [0, \phi_{\max}] \times \mathrm{d} t  \big), \quad j=1,\ldots, p.
\end{equation}
We see that, by construction, $\mathrm{d} \widehat{N}_j(t) \geq  \mathrm{d}  {N}^{(i)}_j(t)$ for any $i$, and thus $\mathrm{d} \widehat{N}_j(t) \geq \lim_{n\to \infty}  \mathrm{d}  {N}^{(i)}_j(t)= \mathrm{d} {N}_j(t)$.
It is then follows that, for any $n$ and $A \in \mathcal{B}(\mathbb{R})$,
\begin{equation}\label{eqn::thinning_tail}
\mathbb{P} \big(N_j\big(A\big)  \geq n \big) \leq   \mathbb{P} \big(\widehat{N}_j\big(A  \big)  \geq n \big).
\end{equation}
Now, by definition $\widehat{N}_j (A )$ is a Poisson random variable.
Then, by Lemma~\ref{lmm::poissontail}, the tail probability in \eqref{eqn::thinning_tail} decreases exponentially fast in $n$.

(ii) Suppose that Assumption~\ref{asmp::bounded}b holds. Recall that we represent the Hawkes process $\bm{N}$ using an iterative construction and a dominating homogeneous Poisson process $\bm{N}^{(0)}$ in Equations~ \ref{eqn::iterative_initial_main} -- \ref{eqn::iterative_construction_main}. Here we construct another process $\widehat{\bm{N}}$ in a similar manner but the intensity of the process $\widehat{\bm{N}}$ takes the form
\begin{equation}\label{eqn::HP_intensity_dom}
\hat{\lambda}_{j}(t)= \phi_j \big(\mu_{j} \big) + \sum_{k=1}^p \big(\hat{\omega}_{k,j} \ast  \mathrm{d} {N}_k \big) (t), \;\;\; j=1,\ldots,p,
\end{equation}
where $\hat{\omega}_{j,k}(\Delta)\equiv|\omega_{j,k}(\Delta)|$.
From the Lipschitz condition of $\phi_j$,  $\hat{\lambda}_{j}(t)$ is always larger than $\lambda_j$ given the same history.
From the iterative construction, we can see that the process $\widehat{{N}}^{(i)}_j$ dominates ${N}^{(i)}_j(A)$ for any $i$, which means that the limiting process $\widehat{N}_j$ dominates $N_j$.
As a result, we have ${{N}_j(A)} \leq \widehat{{N}}_j(A)$ for any bounded interval $A$. For   ${\widehat{N}_j(A)}$, we know from Proposition~2.1 in \cite{reynaud2007} that $P(\widehat{{N}}_j(A)> n ) \leq \exp( 1- n/K)$, which implies that $P({{N}}_j(A)> n ) \leq \exp( 1- n/K)$. In other words, $N_j(A)$ has an exponential tail of order $1$.
\QEDB

\end{document}